\documentclass[nonacm=true,acmsmall]{acmart}
\usepackage{cleveref}
\usepackage{stmaryrd}
\usepackage{alltt}
\usepackage{graphicx}
\graphicspath{{./figures/}{../figures/}}

\usepackage{semantic}

\usepackage{xcolor}
\definecolor{egreen}{rgb}{0.31, 0.78, 0.47}
\definecolor{ate}{rgb}{0.58, 0.0, 0.83}
\newtoggle{revision}
\toggletrue{revision}

\iftoggle{revision}{
    \newcommand{\rjs}[1]{\textit{\textcolor{red}{[rjs]: #1}}}
    \newcommand{\rnr}[1]{\textit{\textcolor{blue}{[Rand]: #1}}}
    \newcommand{\obo}[1]{\textit{\textcolor{egreen}{[Seun]: #1}}}
    \newcommand{\oboc}[1]{\textit{\textcolor{egreen}{#1}}}
    \newcommand{\ate}[1]{\textit{\textcolor{ate}{[Aidan]: #1}}}
}{
    \newcommand{\rjs}[1]{}
    \newcommand{\rnr}[1]{}
    \newcommand{\obo}[1]{}
    \newcommand{\oboc}[1]{}
    \newcommand{\ate}[1]{}
}

\usepackage{bcprules}
\usepackage{backnaur}
\usepackage{pgfplots}
\usepackage{tikz}
\usepackage{multirow}

\usepackage{algorithm}
\usepackage[noend]{algpseudocode}

\usepackage[frame,line,arrow,matrix,tips]{xy}	

\usepackage[T1]{fontenc}
\usepackage{textcomp}
\usepackage[scaled]{beramono}

\usepackage{listings}

\definecolor{codegreen}{rgb}{0,0.6,0}
\definecolor{codegray}{rgb}{0.5,0.5,0.5}
\definecolor{codepurple}{rgb}{0.58,0,0.82}
\definecolor{backcolour}{rgb}{0.95,0.95,0.92}

\definecolor{codegreen}{rgb}{0,0.6,0}
\definecolor{codegray}{rgb}{0.5,0.5,0.5}
\definecolor{codepurple}{rgb}{0.58,0,0.82}
\definecolor{backcolour}{rgb}{0.95,0.95,0.92}

\lstdefinestyle{myStyle2}{            
    backgroundcolor=\color{backcolour},   
    commentstyle=\color{codegreen},
    keywordstyle=\color{magenta},
    numberstyle=\tiny\color{codegray},
    stringstyle=\color{codepurple},
    basicstyle=\small\ttfamily,
  breakatwhitespace=false,
  breaklines=true,
  captionpos=b,
  keepspaces=true,
  numbers=none,              
    numbersep=5pt,                  
    showspaces=false,                
    showstringspaces=false,
    showtabs=false,                  
    tabsize=2
}

\lstdefinestyle{myStyle}{
    belowcaptionskip=1\baselineskip,
    commentstyle=\color{codegreen},
    breaklines=true,
    frame=none,
    numbers=none,
    basicstyle=\footnotesize\ttfamily,
    keywordstyle=\bfseries\color{green!40!black},
    commentstyle=\itshape\color{purple!40!black},
    backgroundcolor=\color{gray!10!white},
    stringstyle=\color{codepurple},
    float=ht,
    frame=tb,
}
\definecolor{mintedbg}{rgb}{0.95,0.95,0.95}

\CompilePrefix{xygui-}

\def\w{\ar@{-}[l]}
\def\W{\ar@{=}[l]}

\def\A#1{\save []="#1" \restore}

\def\op#1{*+[F]{\rule[-0.2ex]{0ex}{2.1ex}#1}}	
\def\b{*={\bullet}}

\def\N{*-{}\W}
\def\n{*-{}\w}

\def\>{\rangle}
\def\<{\langle}

\def\meter{*+[]{\put(-4,0){\includegraphics[scale=.5]{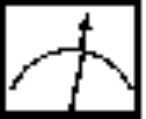}}\hspace{14pt}}%
		\ar@{-}[l]}

\def\qv#1#2{*+{\rule[-0.2ex]{0ex}{2.1ex}|#1\>=|#2\>}}
	
\def\gspace#1{*+{\rule[-0.2ex]{0ex}{2.1ex}%
	\setbox\sbox=\hbox{$#1$}%
	\hspace*{\wd\sbox}}}
	
 \usepackage{xcolor}

\newcommand{\bra}[1]{\langle#1|}
\newcommand{\ket}[1]{|#1\rangle}

\renewcommand{\bnf}{\ensuremath{\mathrel{::=}}}

\acmJournal{PACMPL}
\acmVolume{1}
\acmNumber{OOPSLA} 
\acmArticle{1}
\acmYear{2018}
\acmMonth{1}
\acmDOI{} 
\startPage{1}

\setcopyright{none}

\usepackage{booktabs}   
\usepackage{subcaption} 

\begin{document}

\title[MCBeth: A Measurement Based Quantum Programming Language]{MCBeth: A Measurement Based\\ Quantum Programming Language}         
                                        


\author{Aidan Evans}
\authornote{Contributed equally}          
\affiliation{
  \institution{Yale University}            
}
\email{aidan.evans@yale.edu}          

\author{Seun Omonije}
\authornotemark[1]          
\affiliation{
  \institution{Yale University}           
}
\email{seun.omonije@yale.edu}         

\author{Robert Soul\'{e}}
\affiliation{
  \institution{Yale University}           
}
\email{robert.soule@yale.edu}         

\author{Robert Rand}
\affiliation{
  \institution{University of Chicago}           
}
\email{rand@uchicago.edu}         



\newcommand{\MCL}{MCBeth}

\begin{abstract}
Gate-based quantum programming languages are ubiquitous but measurement-based languages primarily exist only on paper. This work introduces \MCL{}, a quantum programming language which allows programmers to directly represent, program, and simulate measurement-based and cluster state computation by building upon the measurement calculus. While \MCL{} programs are meant to be executed directly on hardware, to take advantage of current machines we also provide a compiler to gate-based instructions. We argue that there are clear advantages to measurement-based quantum computation compared to gate-based when it comes to implementing common quantum algorithms and distributed quantum computation.
\end{abstract}



\maketitle

\section{Introduction}


The quantum computing field has seen tremendous growth in recent years, as new developments in academia and industry have made complex hardware systems a reality. Quantum algorithm design has seen similar progress due to the accessibility of these hardware systems to researchers and the public. As algorithms grow more and more complex, the tools to develop these algorithms need to grow with them. 

Almost all current tools and algorithms, however, use one underlying programming ``language" for reasoning about quantum computation: quantum circuits. The quantum circuit model is a gate-based approach where quantum logic gates are applied to the input, similar to the boolean circuits and logic gates used in classical computation. 
While this gate-based model has widespread use in the quantum community, it constrains our ability to reason about quantum computation as we will discuss below. Thus, current work in quantum computation lacks the benefits that a diverse set of models and languages provide and
forces us to adopt a specific, gate-based intuition in developing quantum programs.

In the case of \textit{classical} logic gates, there are physical gates in the hardware which correspond to each logic gate, and it sometimes makes sense to build programming abstractions around these physical gates. 
In contrast, for \textit{quantum} logic gates, there are no physical quantum gates or circuits: the most common quantum architectures feature arrays of isolated quantum bits (qubits) which are acted on by laser pulses. 
Because of this, quantum circuits provide a poor abstraction for the programmer: by interleaving single-qubit gates, entangling gates, and measurement, they obscure where the power of quantum computation comes from. For example, the most common quantum gate set, the Clifford set, isn't even universal for quantum computation~\cite{gottesman1998}.

The gate-based quantum model was not created to be a programming language in the first place; it was created to demonstrate and reason about quantum computers as a model of computation, not for the practical design of advanced algorithms and hardware systems. Like Turing machines, the gate-based model has proven effective as a theoretical model of computation. Neither our current classical hardware systems nor programming languages, however, are based on Turing machines. Analogously, the gate-based model should not be considered the end-all and be-all choice for reasoning about quantum computation. Other models and other languages may prove more fruitful in the advancement of quantum computation in different domains.


Where the exact power of quantum computation lies is difficult to know. There's not a clear delineation of where the power of quantum computation comes from when looking at gate-based quantum circuits. An alternative approach is measurement-based quantum computation (MBQC), also known as one-way quantum computation.  Many algorithms have been built around the measurement-based quantum computer\cite{Raussendorf_2003}\cite{simons-experiment}\cite{bv-dj-tame}\cite{deutschjozsa2010} despite it being a less known approach. Notably, the measurement calculus\cite{danos} is a general mathematical model for MBQC that provides a foundation for practical measurement-based quantum programming.

The measurement calculus does a better job at conveying the unique power of quantum computation in its structure than gate-based models because measurement-based models 
quantify the amount of entanglement needed at the beginning of the program. In other words, we can glean quantifiable information about a unique property of quantum computation simply by glancing at the program; the operations available also correspond to easily understandable quantum operations, rather than arbitrary gates. Moreover, as we will show, certain programs such as quantum teleportation are much easier to express in the measurement-based model compared to the gate-based model. 

While there exists a plethora of software tools available for dealing with gate-based models, few software tools currently exist for dealing with measurement-based models -- despite promising experimental results of running measurement-based algorithms on real quantum computers~\cite{simons-experiment}. At the moment, all the tools available for effectively working with measurement-based quantum algorithms are primarily theoretical frameworks, like the measurement calculus. The measurement calculus has proven to be a powerful tool for reasoning about MBQC algorithms but there are currently next to no ways to effectively create and simulate programs written in this language. 

This work presents \MCL{}, a programming language designed to bring the measurement calculus out of the theoretical and into practice, allowing for the easy development of measurement-based quantum algorithms. This language is one of the first to provide programmers with the ability to explore, create, run, test, and export measurement-based quantum algorithms. We provide a library containing tools for creating \MCL{} programs via lower-level primitive commands and higher-level commands which can be compiled down to the primitive commands. We provide utility functions for converting the lower-level measurement calculus commands to a standard form and checking program validity. We provide functions for the weak and strong simulation of \MCL{} programs and implementations of eight quantum algorithms written in \MCL{}. Additionally, we provide a translation framework from \MCL{} to other quantum programming frameworks such as OpenQASM~\cite{cross2017open} and Cirq~\cite{cirq}, allowing for cross-compatibility between our language and other gate-based languages and giving programmers the ability to test their programs on a real, public quantum devices.

One of the key benefits of \MCL{} is that we can distribute computation. Since \MCL{} is measurement-based, the standardization procedure allows all two-qubit, entangling operations to be completed at the start of the computation. This allows for the distribution of entangled qubits across a network of quantum computers at the start of the computation. Following the initial distribution of qubits, the computation can proceed on quantum computers which only need to handle a small number of qubits. Furthermore, this allows for the parallelization of \MCL{} programs because certain single-qubit operations throughout the computation can be performed simultaneously on separate computers in the network. We provide functions for constructing distributed \MCL{} programs and converting non-distributed \MCL{} programs into a distributed form. We also provide functions for the pseudo-simulation\footnote{We say ``\textit{pseudo}-simulation'' and not simply ``simulation'' because we cannot always simulate the distributed programs accurately on classical computers. In practice, quantum entanglement allows for the computation to proceed distributively but we cannot classically simulate this entanglement accurately without giving up parallelization in the simulation.}
of these distributed programs.

In this paper, in Section \ref{sec:back} we first provide a background on quantum computing from both a gate-based and measurement-based perspective. We then introduce MCBeth by walking through an example of creating a MCBeth program in Section \ref{sec:byexam} and providing a formal syntax and denotational semantics in Section \ref{sec:langdes}. After formally introducing MCBeth, we provide additional examples of constructing advanced MCBeth programs in Section \ref{sec:programmingMCB}. In Sections \ref{sec:sim} and \ref{sec:arch}, we then turn to discussing the simulation and execution of MCBeth programs in practice, including on MBQC-specific architectures. Finally, in Section \ref{sec:eval} we provide an overall evaluation of MCBeth and in Section \ref{sec:relatedwork} we discuss related work.

\section{Background}\label{sec:back}
    We now provide a brief introduction to quantum computation. We first provide a high-level overview of quantum computation, followed by an introduction to gate-based and measurement-based quantum computation.

  \subsection{Quantum Computing}
      Quantum computing takes advantage of the property that fundamental particles can occupy multiple points in space simultaneously so they can exist in multiple states at the same time.
      Most quantum architectures support two stable states but some may support more; for example, superconducting circuits can easily support three stable states~\cite{gokhale2019}. A discretized piece of quantum information is called a \textit{qudit}. If a quantum architecture supports $n$ stable states per qudit, then the quantum state can be represented as an $n$-dimensional vector which is a linear combination of $n$ \textit{basis states}. A set of basis states is any set of $n$ $n$-dimensional orthogonal vectors. When a state is is not equal to one of the basis states, it is said to be in a \textit{superposition} of states.
      
      Just as classical computation represents information using bits, most quantum computers use \emph{qubits}, which have two stable states. The most common basis state for computation with qubits is the computational bases, $\ket{0}$ and $\ket{1}$. The remainder of this paper assumes the use of qubits to represent quantum information.
      
      A quantum state is \textit{entangled} when it cannot be described in terms of separate quantum states. In other words, the overall state of the entire system can only be described when taking into account multiple particles; each particle in the system cannot be described individually. Mathematically, separate states can be combined into one system using the \textit{tensor product}; we call this product the \textit{state vector} of the system. Hence, we call a state entangled when it cannot be decomposed into the tensor product of each individual qubit's state. 
      
      The act of measurement is an operation on a quantum system that causes the quantum state being measured to collapse to the basis the measurement is performed in. Measurement is referred to as a \textit{destructive operation} because it cannot be reversed; once performed, we cannot recover the pre-measurement quantum state.
      
      Overall, the believed advantage of quantum computation over classical computation comes from superposition and entanglement. With superposition, quantum computation has the ability to have pieces of information represented as one of an exponential number of states. With entanglement, operations on one piece of information can automatically affect others.

    \subsection{Gate-Based Quantum Computing}
    
      In the standard gate-based model, \emph{quantum gates} map one quantum state to another. \emph{Unitary gates} are represented by a unitary matrix, $U$. Because their matrices are unitary, i.e., their adjoints are their inverses, they are \textit{reversible}, which means that applying gate $U$ to a qubit can be undone by applying $U^{-1}$. Gates may act on a single qubit or multiple. An example of a commonly used single qubit gate is the \textit{Hadamard gate} ($H$). Other important single-qubit gates are the Pauli operators: $X$, $Y$, and $Z$; each of these rotate a qubit's state $180^{\circ}$ around the $x$-, $y$-, or $z$-axis when the qubit is represented in three-dimensional space. An important multi-qubit gate is the \textit{controlled-U} gate (sometimes written $CU$), where $U$ is a single qubit gate -- e.g., $CX$ and $CZ$ in the cases of controlled-X and controlled-Z gates. In a controlled-U gate, there's a ``control'' and ``target'' qubit. The operator $U$ is only applied to the target qubit if the control qubit equals $\ket{1}$. The controlled-X gate is typically referred to as the controlled-NOT gate or \textit{CNOT}. The controlled-Z gate will have an important role in later sections of this paper. Typically, a quantum circuit first consists of a set of qubits prepared in some initial state. Then, quantum gates are sequentially applied to the qubits. Finally, the qubits are measured with respect to a basis.

      \textit{Measurement} operations collapse the state of a specified qubit onto a certain basis; typically, in the gate-based model, measurement collapses the state to either $\ket{0}$ or $\ket{1}$. In other words, measurement projects the quantum state of a qubit onto one of the computational bases. Which state a qubit collapses to is probabilistic; specifically, given the state, $\ket{\psi} = \alpha \ket{0} + \beta \ket{1}$ of a qubit, $\ket{\psi}$ will collapse to $\ket{0}$ with probability $|\alpha|^2$, and $\ket{1}$ with probability $|\beta|^2$. 
      
      Because the gate-based model freely interleaves gates which could introduce entanglement, apply single-qubit gates, or perform measurement, it's not clear what roles entanglement and measurement play in a computation. It's also hard to see by glancing at a circuit how much entanglement is generated. Similarly, it's not clear what role measurement plays besides collapsing a state to one of the basis states. The gate-based model obscures the quantum operations being performed, which causes the gate-based model to hide where the power of quantum computation occurs and removes any intuition on how to write effective quantum algorithms.

    We should further note that the gate set typically used to teach gate-based quantum computing ($H$, $X$, $Z$, and $CNOT$ above) isn't even universal for classical computing, let alone quantum computing. In order to approximate arbitrary quantum computations, the gate-based model will frequently add the $T$ gate (obtaining the well-known ``Clifford+T'' gate set); furthermore, to precisely implement any quantum computation, it will tend to also use a fine-grained rotation gate like $R_z(\theta)$, which rotates the qubit around the $z$-axis by $\theta$ degrees.

      
  \subsection{Measurement-Based Quantum Computing}
  
    Measurement-Based Quantum Computing (MBQC) is a promising alternative to the gate-based approach. One important model of MBQC is the \textit{cluster state model} \cite{Nielsen_2006}. The cluster state model typically involves (1) the creation of a collection of entangled qubits called a \textit{cluster state}, (2) the measurement of qubits in different bases, and, in some variations, (3) the application of X and Z \textit{corrections} using the Pauli operators.  The cluster state of the computation is typically obtained by performing a controlled-Z gate between qubits to entangle them. Because the entanglement of the cluster can only decrease as the computation progresses, the cluster state model is often referred to as \textit{one-way quantum computation}. Additionally, the basis of each measurement and whether corrections are performed may be dependent on the outcome of earlier measurements performed during the computation.
    In measurement-based computation, this ordering of entanglement, measurement, and then correction operations is enforced. 
    

    MBQC's forced ordering of operations clearly distinguishes computation into three phases which highlight the true role that entanglement and measurement have in quantum programs, unlike the gate-based model. Because the ordering requires all entanglement to be realized upfront -- instead of buried in the computation -- this shows how entanglement is really a resource for a quantum computation. An initial amount of entanglement is provided upfront in the form of an entangled cluster and the computation proceeds by using measurement to continually remove qubits from the cluster -- thus, disentangling the cluster. This also highlights how the computation itself is primarily performed by the measurement operations. Thus, with MBQC, we have a clear picture of what quantum operation the computation itself primarily depends on; while in the gate-based model, every step of the program (i.e. every gate) is just said to simply perform a computation.
    
    
    
    Computation via cluster states also allows for diagrammatic program construction. The cluster state created after the entanglement phase can be represented as a graph where nodes represent qubits and edges represent entanglement between qubits. This diagram has the ability to capture the computation and, therefore, we can construct and understand programs purely in terms of their respective clusters. We discuss this representation in Section \ref{sec:byexam}.
    
    
    The \textit{measurement calculus} is a language for the cluster state model formalized by Danos et al \cite{danos}. It provides a clear syntax and semantics to the cluster state operations along with rewriting rules for simplifying sequences of commands. The measurement calculus, like the cluster state model in general, is universal for quantum computation; therefore, it can compute anything that the standard gate-based model can. Although there have been several papers on the measurement calculus, to date, there are next to no practical languages which implement it. This makes MCBeth one of the first implementations of the measurement calculus and one of the first implementations of a practical language for MBQC.

    
\section{MCBeth by Example: Teleportation}\label{sec:byexam}
    We now introduce MCBeth by walking through the implementation of the quantum teleportation algorithm. To highlight how MCBeth differs from gate-based abstractions, we provide implementations of teleportation using both the gate-based model and the measurement calculus.

    \begin{figure}[b]
      \includegraphics[scale=0.8]{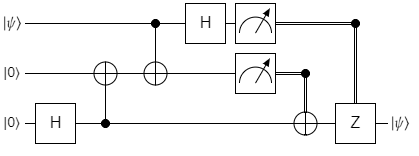}
      \begin{center}
          \textbf{Traditional Teleportation Circuit}
      \end{center}
      \vspace{5mm}
      
      \begin{minipage}{0.48\textwidth}
        \centering
        \begin{tikzpicture}  
          [scale=.9,auto=left,every node/.style={circle,fill=blue!20}]
            
          \node (a1) at (1,1) {$q_0$};  
          \node (a2) at (3,1) {$q_1$};  
          \node (a3) at (5,1) {$q_2$};
          
          \draw (a1) -- (a2); 
          \draw (a2) -- (a3);
    
        \end{tikzpicture}
        \vspace{4mm}
          \begin{center}
            $J(\alpha)(q_0, q_1); J(\beta)(q_1,q_2)$ where $\alpha,\beta = 0$
          \end{center}
          \vspace{2mm}
      \end{minipage}\hfill
      \begin{minipage}{0.48\textwidth}
        {
    
          \def\gAxA{\op{R_z(-\alpha)}\w\A{gAxA}}
          \def\gBxA{\op{R_x(-\beta)}\w\A{gBxA}}
    
    
          \def\bA{\qv{q_{2}}{q_0}}
    
    
          \xymatrix@R=5pt@C=10pt{
              \bA & \gAxA &\gBxA &\n
          %
          %
          }
        }
        \vspace{5mm}
        \begin{flushleft}
          $\ket{q_2} = R_x(0)R_z(0)\ket{q_0} = \ket{q_0}$
        \end{flushleft}
        \vspace{2mm}
      \end{minipage}
      
        \begin{center}
            \textbf{Teleportation Cluster and Commands}
            \hspace{10mm}
          \textbf{Cluster Effect on Original Quantum State}
        \end{center}
        \caption{Teleportation Circuit contra Cluster}
      \label{fig:teleportation}
    \end{figure}

    Teleportation is fundamental to any form of distributed quantum computing and quantum networking. Teleportation is important because it provides a way to transfer a quantum state between two qubits without using a quantum channel or violating the \textit{no-cloning theorem}. The no-cloning theorem states that it's not possible to \textit{copy} a quantum state from one qubit to another. Teleportation, however, allows us to at least \textit{transfer} the quantum state to another qubit via entanglement and the sending of information over classical channels.
    
    \paragraph{Gate-based Teleportation}
    Figure~\ref{fig:teleportation} depicts the traditional quantum circuit for this process. The first four gates -- the two Hadamard and CNOT gates -- entangle the qubits together. Measurement is then performed on the first and second qubits. Finally, X and Z gates are applied to the last qubit based on the measurement outcomes of the first and second qubits. One may not find it clear, however, how the qubits of this circuit actually affect one another and why the state ends up being transferred from the first qubit to the last qubit. The measurement calculus, on the other hand, does help us reason about how the qubits interact with each other and about the overall computation in general.
    
    \paragraph{Using Clusters to Construct Measurement-based Programs}
    We can start by approaching the problem diagrammatically. We want to find a \textit{cluster} that will render the desired effect on the input qubits' states during computation; in this case, we wish for the our initial qubit's state to be transferred identically to some other qubit. We will ultimately use the cluster depicted in Figure~\ref{fig:teleportation}.
    
    Each distinct cluster has an associated program and can be interpreted as performing a computation starting on the leftmost qubits in a cluster and propagating the computation through the cluster until the final state is remaining in the rightmost qubits of the cluster. This relationship between the leftmost initial qubits and rightmost final qubits can be mapped to a quantum circuit acting on $n$ qubits, where $n$ is the number of rows in our cluster. We display several example clusters with their corresponding circuits and commands collectively in Section \ref{app:clusters} of the Appendix. Reading the graphs of each cluster from left to right, nodes in the first column represent input qubits and nodes in the last column represent output qubits. 
    
    Edges between nodes in the same row represent the application of the command $J(\alpha)(q_i, q_j)$ in the measurement calculus. The $J(\alpha)(q_i, q_j)$ command is a high-level command which compiles down to several lower-level primitive commands; specifically, it represents the process of entangling qubits $q_i$ and $q_j$, measuring $q_i$, and then correcting $q_j$ as necessary, where $q_i$ is the left qubit, $q_j$ is the right qubit, and $\alpha$ is the \textit{angle of measurement} for $q_i$. We explain this further below and provide an explicit semantics for $J$ in Section \ref{sec:langdes}.
    
    Edges between nodes in different rows represent the application of the $CZ(q_i, q_j)$ command, which just entangles $q_i$ and $q_j$ -- as seen in Figure \ref{fig:djcluster}.
    
    We can also concatenate two clusters together by merging the nodes in the last column of one cluster with the nodes in the first column of another cluster.
    
    \paragraph{Writing the Measurement-based Teleportation Program in MCBeth}
    For our purposes, we want the three qubit cluster depicted in Figure~\ref{fig:teleportation}. Here we have three qubits positioned in a horizontal line with entanglement between $q_0$ and $q_1$ and between $q_1$ and $q_2$. Computation then proceeds by measuring $q_0$ in a certain basis, impacting $q_1$'s state, and then measuring $q_1$, impacting $q_2$'s state. Specifically, we get the following program written in the measurement calculus:\[
        J(\alpha)(q_0,q_1); J(\beta)(q_1,q_2)
    \]
     For teleportation, because we wish for the final state to equal the initial state, we set $\alpha,\beta=0$ and now get \[
        J(0)(q_0, q_1); J(0)(q_1, q_2)
    \]
    Figure~\ref{fig:teleportation} shows the effect this will have on the original state of qubit $q_0$. Note that this corresponds to performing an $R_z$ rotation of $0$ and $R_x$ rotation of $0$; i.e., the state is not changed yet because the computation was propagated through the clusters, we end up with our state from $q_0$ in $q_2$ as desired.
    
    In MCBeth, we would represent these commands using the following program:
    \begin{lstlisting}
    Input(0);
    PrepList([1, 2]);
    J(0.0, 0, 1);
    J(0.0, 1, 2);\end{lstlisting}
    
    Here, we have a list of commands. The ``\texttt{J}(0.0, 0, 1);'' command is the command for $J(0)(q_0, q_1)$.
    
    This program can then be both decomposed into lower-level primitive commands. We now describe what these primitive commands are and how the decomposition process is performed.
    
    \paragraph{Decomposing into Primitive Commands}
    The commands \[
        J(0)(0, 1); J(0)(1, 2)
    \] helps us make sense of the circuit in Figure~\ref{fig:teleportation} because in the measurement calculus, high-level commands such as $J$ are actually slight abstractions of lower-level quantum operations we call \textit{primitive commands}. These include, preparation ($N_i$), entanglement ($E_{ij}$), measurement ($M_i^\alpha$), and Pauli-X and Pauli-Z corrections ($X_i$ and $Z_i$). We now discuss these commands individually and their associated MCBeth counterparts; further detail of the MCBeth language is provided in Section \ref{sec:langdes}.

    The preparation command, $N_i$, initializes qubit $i$ to the $\ket{+}$ state. The $\ket{+}$ and $\ket{{-}}$ states are two common states used in quantum computation:
    \begin{align*}
        \ket{+} &= \frac{1}{\sqrt{2}}(\ket{0} + \ket{1})\\
        \ket{{-}} &= \frac{1}{\sqrt{2}}(\ket{0} - \ket{1})
    \end{align*}
    In MCBeth, the initialization command is represented by the command  \texttt{Prep($q$)} where $q$ is a qubit; thus, to initialize a qubit 0, we write \texttt{Prep(0)}. For convenience, we also provide a command which allows one to initialize a list of qubits at once: \texttt{PrepList([0; 1; 4])}. This initializes qubits 0, 1, and 4 to the $\ket{+}$ state.
    
    Since some qubits in a program may be passed in as input as opposed to initialized, we signify this distinction by having an input command: \texttt{Input($q$)}. During simulation, a mapping of qubits to states can then be passed into the simulator which will set each input qubit $q$ to it's desired state at the start of the simulation.
    
    All qubits must either be declared using the \texttt{Prep} or \texttt{Input} commands at the start of the program.
      
    The measurement calculus's entanglement command, $E_{ij}$, entangles qubits $i$ and $j$ by performing a controlled-Z gate between them with $i$ the control and $j$ the target. In MCBeth, we use the command \texttt{Entangle($q_1$, $q_2$)}.
    
    The measurement command, $M_i^\alpha$, measures qubit $i$ with the following basis states:
    \begin{align*}
        \ket{+_\alpha} &= \frac{1}{\sqrt{2}}(\ket{0} + e^{i\alpha}\ket{1})\\
        \ket{{-}_\alpha} &= \frac{1}{\sqrt{2}}(\ket{0} - e^{i\alpha}\ket{1})
    \end{align*}
    where $\alpha \in [0, 2\pi]$; $\alpha$ is the \textit{the angle of measurement}. Notice the when $\alpha = 0$, measurement is performed with the $\ket{+}$ and $\ket{-}$ basis states. In MCBeth, we write \texttt{Measure}($\alpha$, $q$, $signals_s$, $signals_t$) where $\alpha$ is a float and $signals$ are lists of qubits -- signals are explained further below.
    
    The correction commands, $X_i$ and $Z_i$, apply the Pauli $X$ or $Z$ gates, respectively, to qubit $i$. In MCBeth we write, $\texttt{XCorrect}(q, signals)$ and $\texttt{ZCorrect}(q, signals)$.
    
    The measurement and correction commands may depend on the results of past measurements performed. The outcome of a measurement on qubit $i$ is denoted as $s_i$ where
    \begin{align*}
        s_i = \begin{cases}
                0 &\text{if the qubit collapsed to } \ket{+_\alpha}\\
                1 &\text{if the qubit collapsed to } \ket{ -_\alpha}
              \end{cases}
    \end{align*}
    Dependent commands use \textit{signals} to determine the operation performed. Signals are the exclusive-or of measurement outcomes. In other words, signal $s = \left(\oplus_{i \in Q}{s_i}\right)$ where $Q$ is the set of qubits the commands depends on. $Q$ is referred to as the \textit{domain} of a signal.
    
    Dependent corrections are written $X^s_i$ and $Z^s_i$. The operation is performed if $s = 1$ and not performed if $s = 0$. For example, in the case of $X^s_i$, $X^1_i = X_i$ and $X^0_i = I_i$ where $I$ is simply the identity matrix.
      
    Dependent measurements are written as 
    \begin{align*}
        {}^t[M_i^\alpha]^s = M_i^{(-1)^s\alpha + t\pi}
    \end{align*}
    where $t$ and $s$ are signals. 
    In MCBeth, these signals are included in the measurement and correction commands as lists of qubits.

    A measurement calculus program then consist of a finite sequence of commands and three sets of qubits: $V$, $I$, and $O$. $V$ is the computational space, i.e., the set of all qubits used in the computation. $I$ is the set of input qubits and $O$ is the set of output qubits. Moreover, in this paper, a sequence of commands, $A_1 ; \dots ; A_n$ is read from left to right.\footnote{In Danos et al.'s measurement calculus, commands are written from right to left so mathematical operations, i.e., matrix multiplication, is easier expressed. Since we are designing a language for programming, we choose the standard representation for sequencing.} Note that $I \cup O \subseteq V$ and the sets $I$ and $O$ may share the same qubits -- i.e., $I \cap O$ does not necessarily equal $\varnothing$.

    Programs must abide by the following constraints:
    \begin{enumerate}
        \item no command depends on an outcome not yet measured
        \item no command acts on a qubit already measured
        \item no command acts on a qubit not yet prepared, unless it is an input qubit
        \item a qubit $i$ is measured by some $M_i^\alpha$ if and only if $i$ is not an output qubit
    \end{enumerate}
    The constraints ensure that a program is well-formed so that the execution of the commands makes sense. 

    Moreover, the following set of high-level commands are universal for quantum computation; i.e., any quantum algorithm can be implemented using only these commands:
    \begin{align*}
        C Z(i, j) &:= E_{ij}\\
        J(\alpha)(i, j) &:= E_{ij}; M^{-\alpha}_i; X^{s_i}_j
    \end{align*}
    where $V = \{i, j\}$ for both commands; for $CZ$, $I = O = \{i,j\}$; for $J(\alpha)(i, j)$, $I = \{i\}$ and $O = \{j\}$.

    \paragraph{Standardizing Command Order}

    We also desire to have all entanglements happen before measurement; this process is known as \textit{standardization}. Standardizing a program takes an arbitrary, well-formed sequence of commands and rewrites the program to put commands in the following order: preparations, entanglement, measurement, corrections. The program is rewritten using a set of rewriting rules laid out by Danos et al. Once standardized, the program is simplified and in the proper format resembling cluster state computation. This is easily done in MCBeth by passing a list of commands into the \texttt{standardization} function; given a list of commands it will return an equivalent standardized list of commands.
    
    Importantly, the rewriting rules preserve the semantics of the program; therefore, the standardized program will result in the same computation as the original, non-standardized one.

    The rewriting rules from \cite{danos} are as follows:
    \begin{align*}
        X^s_i; E_{ij} &\Rightarrow E_{ij}; Z^s_j; X^s_i&\\
        X^s_j; E_{ij} &\Rightarrow E_{ij}; Z^s_i; X^s_j&\\
        Z^s_i; E_{ij} &\Rightarrow E_{ij}; Z^s_i&\\
        Z^s_j; E_{ij} &\Rightarrow E_{ij}; Z^s_j&\\
        X^r_i; {}^t[M^\alpha_i]^s &\Rightarrow {}^t[M^\alpha_i]^{s+r}&\\
       \hspace{120pt} Z^r_i; {}^t[M^\alpha_i]^s &\Rightarrow {}^{r+t}[M^\alpha_i]^s&\\
        A_{\vec{k}}; E_{ij} &\Rightarrow E_{ij}; A_{\vec{k}} &\text{where $A$ is not an entanglement}\\
        X^s_i; A_{\vec{k}} &\Rightarrow A_{\vec{k}}; X^s_i &\text{where $A$ is not a correction}\\
        Z^s_i; A_{\vec{k}} &\Rightarrow A_{\vec{k}}; Z^s_i &\text{where $A$ is not a correction}
      \end{align*}%
    where $\vec{k}$ is a set of qubits the command is acting on.

    Returning to our example of teleportation, we can decompose our program
    \begin{align*}
      J(0)(0, 1); J(0)(1, 2)
    \end{align*}
    into 
    \begin{align*}
      E_{01}; M_0; X_1^{s_0};
      E_{12}; M_1; X_2^{s_1}
    \end{align*}
    
    Here, $V=\{0, 1, 2\}$, $I=\{0\}$, and $O=\{2\}$; in other words, the quantum state of qubit 0 is being transferred to qubit 2 and because 1 and 2 are non-input, they are initialized to the $\ket{+}$ state.

    Then, using the rewriting rules, we can rewrite $X_1^{s_0};E_{12}$ as $E_{12};Z^{s_0}_2;X^{s_0}_1$. Thus, we now have
    \begin{align*}
      E_{01}; M_0; E_{12};Z^{s_0}_2;X^{s_0}_1;M_1; X_2^{s_1}
    \end{align*}
    Then, we can rewrite $X^{s_0}_1;M_1$ as $[M_1]^{s_0}$ to get
    \begin{align*}
      E_{01}; M_0; E_{12}; Z^{s_0}_2;[M_1]^{s_0}; X_2^{s_1}
    \end{align*}
    Finally, applying some final rules to commute the commands, we get a standardized program:
    \begin{align*}
      E_{01}; E_{12}; M_0; [M_1]^{s_0}; Z^{s_0}_2; X_2^{s_1}
    \end{align*}
    Thus, in MCBeth, we would obtain the standardized program shown in Figure \ref{fig:teleprog}.
    
    \begin{figure}[t]
        \centering
        \begin{lstlisting}
        Input(0);
        PrepList([1, 2]);
        Entangle(0, 1);
        Entangle(1, 2);
        Measure(0, 0.0, [], []);
        Measure(1, 0.0, [0], []);
        ZCorrect(2, [0]);
        XCorrect(2, [1]);\end{lstlisting}
        \caption{Teleportation in MCBeth after Standardization}
        \label{fig:teleprog}
    \end{figure}
    Note that standardization now allows us to perform the initial entanglement of the cluster during the same phase of the computation. Furthermore, if we were to create the initial entanglement of the qubits on one computer and then distributed these qubits over a quantum network to two other nodes such that qubits 0 and 1 are on one node and qubit 2 is on another, then we can still transfer the quantum state of qubit 0 to qubit 2 over any distance, as long as we can transfer the signals $s_0$ and $s_1$ over a classical channel. Further application of this distributed architecture is explained in section \ref{sec:distributed}. 
    
    \paragraph{Comparison to the gate-based version}
    
    By using the measurement calculus, we can see how a circuit such as the one in Figure~\ref{fig:teleportation} could be constructed to teleport a quantum state between two qubits. We first constructed a cluster state which allowed us to have the desired effect on our input state -- in this case, no effect. We then constructed a program representing this cluster, decomposed the program into low-level primitive commands, and standardized the program. This process gave us a sequence of quantum operations where entanglement happens first, then measurement, and finally corrections -- similar to that of our circuit in Figure~\ref{fig:teleportation}. 
    
    In the gate-based model, using circuit identities, the $CNOT$ and Hadamard gates can be converted to two $CZ$s. Remember that our entanglement command, $E_{ij}$, is equivalent to a $CZ$; therefore, the $CNOT$ and Hadamard gates entangle the three qubits together. The circuit then measures two qubits and uses this outcome to apply dependant $X$ and $Z$ gates, just like in the measurement calculus. 
    
    In the measurement calculus, however, we had a clear process of how to approach setting up the cluster and the effects it would have on our final qubit; we were easily able to reason about how the state would be transferred among qubits using the $J$ operator. We did not have this ability with the gate-based circuit.
    

\section{Language Design}
\label{sec:langdes}

\subsection{Syntax}
    We now formally introduce our implementation of the measurement calculus: \MCL{}. The syntax of our language is defined in Figure~\ref{fig:syntax}. A \MCL{} program consists of a list of commands separated by semicolons. The supported commands break down into three general categories: input preparation, core execution, and readout. When writing a program, qubits are referred to as natural numbers.
    
    The input preparation phase consists of two types of commands: ``Input'' commands and ``Prep'' commands. ``Input'' allows one to declare which qubits should be passed in as input during the start of the computation while ``Prep'' initializes the qubit to $\ket{+}$. ``InputList'' and ``PrepList'' commands are added to allow for the easy initialization of a list of qubits.

    The main execution phase consists of applying the commands for entanglement, measurement, and Pauli corrections to the quantum system initialized in the input preparation phase. For the applicable commands, signals are passed in as lists of qubits.

    Finally, an optional readout phase happens which performs a final measurement on specified qubits in a custom basis and stores measurement outcomes as output that should be reported at the end of execution. This is done using the \texttt{ReadOut($...$)} command.

    

    \begin{figure}
      \centering
      \small
      \(\begin{array}{r@{\;}c@{\;}l@{\quad}l@{\quad}l}
      q & \in & \mathbb{N} & \textit{Qubit Label} & \\
      \alpha & \in & [0, 2\pi] & \textit{Angle of Measurement}& \\
      s & \in & \mathbb{C}^2 & \textit{Quantum State}&\\
      \hline
      prog & \bnf & [cmd_1, cmd_2, ...]& \text{Program} & \\
      
      cmd & \bnf & \textit{Input}(q) & \text{Declare Input Qubit} & \\
        &  &  \mid \textit{Prep}(q) & \text{Prepare Qubit} & \\
        &  &  \mid \textit{InputList}(qs) & \text{Declare List of Input Qubits} & \\
        &  &  \mid \textit{PrepList}(qs) & \text{Prepare List of Qubits} & \\
        &  &  \mid \textit{Entangle}(q_1, q_2) & \text{Entanglement} & \\
        &  &  \mid \textit{Measure}(q, \alpha, qs_1, qs_2) & \text{Processing Measurement} & \\
        &  &  \mid \textit{XCorrect}(q, qs) & \text{X Correction} & \\
        &  &  \mid \textit{ZCorrect}(q, qs) & \text{Z Correction} & \\
        &  &  \mid \textit{ReadOut}(q, basis) & \text{Read-out Measurement} & \\
        &  &  \mid \textit{J}(\alpha, q_1, q_2) & \text{J High-Level Command} & \\
        &  &  \mid \textit{CZ}(q_1, q_2) & \text{CZ High-Level Command} & \\
      
      qs & \bnf & [q_1, q_2, ...] & \text{List of Qubits}\\

      basis & \bnf & \textit{X} \mid \textit{Y} \mid \textit{Z} & \text{Common Bases}&\\
          & & \mid \textit{FromTuples}(s_1, s_2)& \text{Custom Basis}&\\
          & & \mid \textit{FromAngle}(\alpha)& & 

      \end{array} 
      \)
      \caption{Syntax of \MCL{}}
      \label{fig:syntax}
      
    \end{figure}

\subsection{Denotational Semantics}
Let $P$ be a program with a computation space of $V$, inputs $I$, outputs $O$, and command sequence $A_1;A_2;...;A_n$. Let $H$ denote the Hilbert space characterizing the qubits of our computation. The state of our system is characterized by both a Hilbert space of non-measured qubits and an output map containing whether a measured qubit collapsed to equal 0 or 1. A command essentially defines a map from an input state to one or more possible output states. Thus, our semantics will interpret command expressions as relations from states to states.

To begin, for convenience, we treat all of the initial input and preparation commands as one expression. In other words, say that once standardized, commands $A_1$ through $A_i$ are all the preparation and input commands. For commands $A_1;\dots;A_i$, the function $\mathcal{C} \llbracket A_1 \dots A_i \rrbracket$ will produce the initialized Hilbert space along with an empty outcome map. Thus, \[
    \mathcal{C} \llbracket A_1 \dots A_i \rrbracket = q_0 \otimes q_1 \otimes \dots \otimes q_n, \varnothing
\]
Here, each qubit $q_j$ is in the state it should be after the initialization process so if we had the command $\texttt{Prep}(2)$, then $q_2 = \ket{+}$. The ``$\varnothing$'' denotes our currently empty outcome map. Say our outcome map is $\Gamma$; note that we'll use $\Gamma[0/j]$ to denote that qubit $j$ maps to 0. Also, we'll use function $f$ as a map of each qubit's number to it's position in the current system; since measurement projects the system down a dimension, the qubit measured is fully removed from $H$ and, therefore, qubit ``i'' may not be the $i^{th}$ qubit in $H$ anymore.

Thus, the remaining denotational semantics for the commands are as follows:
\begin{align*}
    &\comp{Entangle(i, j)}(H, \Gamma) = E_{f(i),f(j)} H, \Gamma\\
    &\comp{XCorrect(i, ss)}(H, \Gamma) = X_{f(i)}^{\oplus_{s \in ss}\Gamma(s)}H, \Gamma\\
    &\comp{ZCorrect(i, ss)}(H, \Gamma) = Z_{f(i)}^{\oplus_{s \in ss}\Gamma(s)}H, \Gamma
\end{align*}
\begin{align*}
    &\comp{Measure(i, $\alpha$, ss, ts)}(H, \Gamma) = \begin{cases}
            \bra{+_{(-1)^{\oplus_{s \in ss}\Gamma(s)}\alpha + (\oplus_{t \in ts}\Gamma(t))\pi}}_{f(i)} H, \Gamma [0/i]\\
            \bra{-_{(-1)^{\oplus_{s \in ss}\Gamma(s)}\alpha + (\oplus_{t \in ts}\Gamma(t))\pi}}_{f(i)} H, \Gamma [1/i]
        \end{cases}\\
    &\comp{J($\alpha$, i, j)}(H, \Gamma) = \begin{cases}
            \bra{+_{-\alpha}}_{f(i)}E_{f(i),f(j)} H, \Gamma [0/i]\\
            X_{f(j)}\bra{-_{-\alpha}}_{f(i)}E_{f(i),f(j)} H, \Gamma [1/i]
        \end{cases}\\
    &\comp{CZ(i, j)}(H, \Gamma) = E_{f(i),f(j)} H, \Gamma\\
    &\comp{ReadOut(i, \textit{basis})}(H, \Gamma) = \begin{cases}
            \bra{\psi}_{f(i)} H, \Gamma[0/i]\\
            \bra{\phi}_{f(i)} H, \Gamma[1/i]
        \end{cases}\quad \text{where }\mathcal{S}|[\textit{basis}|] = (\ket{\psi}, \ket{\phi})\\
    &\mathcal{S}|[X|] = (\ket{+}, \ket{ -})\\
    &\mathcal{S}|[Y|] = (\ket{i}, \ket{ -i})\\
    &\mathcal{S}|[Z|] = (\ket{0}, \ket{1})\\
    &\mathcal{S}|[\texttt{FromTuples((a,\ b),\ (c,\ d))}|] = \left(
    \begin{pmatrix} a\\ b \end{pmatrix}, \begin{pmatrix} c\\ d \end{pmatrix}
    \right)\\
    &\mathcal{S}|[\texttt{FromAngle($\alpha$)}|] = (\ket{+_\alpha}, \ket{ -_\alpha})
\end{align*}
    
where

\[E_{i,j} = M_0 \otimes \dots \otimes M_{n-1} + N_0 \otimes\dots\otimes N_{n-1}\] such that $n$ is the total number of qubits in $H$ during this step of the computation, 
\begin{align*}
    M_i = \ket{0}\bra{0},\ N_i = \ket{1}\bra{1},\ N_j =         
    \begin{bmatrix}
        1 & 0\\
        0 & -1 
    \end{bmatrix}, \text{ and all other } M_k, N_k = I_2 =  
    \begin{bmatrix}
        1 & 0\\
        0 & 1 
    \end{bmatrix}
\end{align*}
and
\begin{align*}
    X_i &= M_0 \otimes \dots \otimes M_{n-1} \text{ such that } M_i = \begin{bmatrix}
        0 & 1\\
        1 & 0
    \end{bmatrix} \text{ and all other $M_k = I_2$}\\
    Z_i &= M_0 \otimes \dots \otimes M_{n-1} \text{ such that } M_i = \begin{bmatrix}
        1 & 0\\
        0 & -1
    \end{bmatrix} \text{ and all other $M_k = I_2$}\\
    \bra{\psi}_i &= M_0 \otimes \dots \otimes M_{n-1} \text{ such that } M_i = \bra{\psi} \text{ and all other $M_k = I_2$}\\
\end{align*}

Note that in the case of measurement we get two branches of our computation: one for each case of how the measured qubit could collapse because, for example, measurement could collapse the qubit to $\ket{+_\alpha}$ or $\ket{ -_\alpha}$; therefore, our semantics reflects this by branching the program along one path in which the qubit measured collapses to $\ket{+_\alpha}$ and one for $\ket{ -_\alpha}$.

\section {Programming in MCBeth}\label{sec:programmingMCB}
    Now that we've introduced the formal syntax and semantics for MCBeth, we now provide implementations of additional algorithms in MCBeth beyond the example of teleportation already discussed.

  \subsection{The Deutsch-Jozsa Algorithm}
    Deutsch's Problem~\cite{deutsch} was one of the problems used to reason about the power quantum computing. The question is as follows: given a black box function that implements $f: \{0, 1\} \rightarrow \{0, 1\}$, determine if that function is constant ($f(0) = f(1)$) or balanced ($f(0) \neq f(1)$). A quantum computer can solve this using a single query to the oracle. 
    
    The Deutsch-Jozsa~\cite{dj-main-paper} algorithm generalizes Deutsch's algorithm and was an early example of a quantum algorithm that could solve a problem faster than the best classical algorithm. It is widely considered the starting point for understanding quantum algorithms. This is the case due to the straightforwardness of the problem, and the relative simplicity of implementation. 
    
    The algorithm solves the Deutsch-Jozsa problem, which is stated as follows: provided a quantum oracle which implements a function $f: \{0, 1\}^n \rightarrow \{0, 1\}$ and the knowledge that this function is either constant (that is, every output is either 0 or 1) or balance (half the input domain returns 1, the other half returns 0), determine if $f$ is constant or balanced by using the quantum oracle. 
    Classically, this problem takes 2 queries to the oracle to determine if the function is balanced in the best case, and $2^{n-1} + 1$ queries in the worst case. Using a quantum computer, however, we can be 100\% confident in an answer with just one query to the oracle.
    
    \begin{figure}[!bp]
     \centering
     \begin{minipage}[b]{0.4\textwidth}
          
        \includegraphics[scale=0.8]{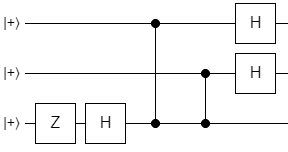}
        \caption{Deutsch-Jozsa Circuit with Balanced Oracle; for constant, simply remove the two CZ gates in the middle.}
        \label{fig:djcirc}
     \end{minipage}
     \hfill
     \begin{minipage}[b]{0.4\textwidth}
        \begin{tikzpicture}  
            [scale=.9,auto=center,every node/.style={circle,fill=blue!20}]
            
            \node (a3) at (1, 2) {$q_3$};  
            \node (a5) at (3, 2) {$q_5$};
            
            \node (a2) at (1,0.5) {$q_2$};
            \node (a4) at (3,0.5) {$q_4$};
            
            \node (a0) at (-1,-1) {$q_0$};
            \node (a1) at (1,-1) {$q_1$};

            \draw (a0) -- (a1);
            \draw (a2) -- (a4);
            \draw (a3) -- (a5);
            
            \draw (a1) -- (a2);
            \draw (a1) to[out=135,in=-135] (a3);
            
            
    \end{tikzpicture}  
    \caption{Deutsch-Jozsa Cluster -- corresponds exactly to the circuit in Figure \ref{fig:djcirc}.}\label{fig:djcluster}
     \end{minipage}
        \vspace{4mm}
        \begin{lstlisting}
        PrepList([0, 1, 2, 3]);
        J(pi, 0, 1);
        CZ(1, 2);
        CZ(1, 3);
        ReadOut(1, Z);
        ReadOut(2, X);
        ReadOut(3, X);\end{lstlisting}
        \caption{Deutsch-Jozsa in MCBeth with Balanced Oracle and Read-out Optimizations}
        \label{fig:djprog}
    
    \end{figure}

    We present a 6-qubit cluster state in Figure \ref{fig:djcluster} handling the balanced oracle for the Deutsch-Jozsa algorithm. This cluster was inspired by prior work in the community \cite{deutschjozsa2010} \cite{bv-dj-tame}. In Figure \ref{fig:djprog}, we provide the \MCL{} program for the problem; we also show here how taking advantage of different measurement bases allows us to remove two qubits from the cluster state ($q_4$ and $q_5$) and achieve the same algorithm. We also compare it to the circuit version displayed in Figure \ref{fig:djcirc} to demonstrate the differences in readability and operation count for programs with the same output. 
  
    The measurement calculus program for this cluster is below, and can be extended to accomodate an additional bit $x$ by adding a $CZ$ operator from qubit 1 to qubit $x$, and a J operator from qubit $x$ to an additional qubit $y$:
    \begin{align*}
      J(\alpha)(q_0,q_1);
      CZ(q_1, q_2);
      CZ(q_1, q_3);
      J(\alpha)(q_2, q_4);
      J(\alpha)(q_3, q_5);
    \end{align*}
    
    As displayed in the \MCL{} program outlined in Figure \ref{fig:djprog}, performing readout measurements on qubits 1, 2, and 3 in the bases shown instead of all in the same base allows us to remove the need for qubits 4 and 5. 
    
    Furthermore, removing the $CZ$ operations and leaving the rest of the cluster intact will implement the constant oracle for the problem.

  

  \subsection{2-bit Grover's Algorithm}
    Grover's algorithm allows one to search for a marked element
    in an unstructured list in $O(\sqrt{n})$ time, where $n$ is 
    the size of the list. It's
    one of the iconic quantum algorithms used to demonstrate
    the generality of quantum speedup. Like the Deutsch-Jozsa algorithm,
    it works by making a call to an oracle function with the
    input being a superposition of states, $\ket{+}$ to be specific.

    Grover's algorithm then works in two parts: (1) using an oracle to mark the desired item in the quantum state and (2) using a diffuser operator to increase this marked item's amplitude and decrease the other items'. Therefore, upon measurement, the item we are searching for will be returned with high-probability. 
    
    The oracle and the diffuser work in tandem in a process called \textit{amplitude amplification}, which reflects the state of the system back and forth over a state $\ket{s'}$, which spans both the state of the desired item and the state of the system. In marking the desired item, the oracle performs the initial reflection, and the diffuser applies a negative phase to that result, amplifying the amplitude of the desired item and lowering the amplitudes of all other options. In small search spaces, one iteration of both the oracle and the diffuser is sufficient; however, as the search space scales, multiple calls to this process are necessary to amplify the state of the winner sufficiently. 
    
    Once the state of the desired item is properly amplified, measurement of the system will return the state with a significantly higher probability than any other state in the system.
    
    A circuit diagram for a 2-bit Grover's algorithm is shown in Figure~\ref{fig:grovercirc}. The first section of the circuit contains a controlled-Z gate between qubits $q_1$ and $q_2$ followed by rotations on each qubit about the Z-axis through an angle of $-\alpha$ and $-\beta$, respectively. This first section represents the oracle and how the oracle is configured is based on the values of $\alpha$ and $\beta$. In other words, we set $\alpha$ and $\beta$ to determine what we are searching for. If, for example, we set $\alpha = 0$ and $\beta = \pi$, then this specifies a search for $1$ on our first input qubit and a $0$ on our second input qubit. The second section, i.e., the rest of the circuit, represents the reflector.
    
    We present two corresponding
    possible cluster states for this circuit based on work by Walther et al. \cite{Walther_2005}: a 6-qubit variation in Figure~\ref{fig:2bitgrover6} and a
    4-qubit variation in Figure \ref{fig:2bitgrover4} which takes advantage of the readout measurement base like we did in the Deutsch-Jozsa program in Figure \ref{fig:djprog}. 
    
    For both variations, the angle of 
    measurement for $q_1$ and $q_2$ determines the oracle
    configuration. An angle of $0$ on qubit $q_i$ specifies a search
    for $1$ on bit $i$ and an angle of $\pi$ specifies 
    a search for $0$ on bit $i$. Then, for the 6-qubit variation,
    qubits $q_3$ and $q_4$ are measured at an angle of $0$
    and puts the output in qubits $q_5$ and $q_6$. Qubits
    $q_5$ and $q_6$ can then be passed as input to another
    algorithm or they can then be read out in the computational 
    basis to obtain the result. For the 4-qubit variation,
    we can instead just perform a readout measurement on 
    $q_3$ and $q_4$ in the $\{\ket{ -}, \ket{+}\}$ basis
    to obtain the result.

    In terms of high-level commands, the 6 qubit variation can be written
    as follows:
    \begin{align*}
      CZ(q_1, q_2);
      J(\alpha)(q_1,q_3);
      J(\alpha)(q_2,q_4);\\
      CZ(q_3, q_4);
      J(\alpha)(q_3,q_5);
      J(\alpha)(q_4,q_6);
    \end{align*}
    followed by readout measurements in the computational basis.

    Similarly, or the 4 qubit variation:
    \begin{align*}
      CZ(q_1, q_2);
      J(\alpha)(q_1,q_3);
      J(\alpha)(q_2,q_4);
      CZ(q_3, q_4);
    \end{align*}
    followed by readout measurements in the $\{\ket{ -}, \ket{+}\}$ basis. Figure \ref{fig:grovprog} displays the corresponding MCBeth program for the 4-qubit variation.

  \begin{figure}
    \begin{minipage}{0.18\textwidth}
    \end{minipage}\hfill
    \begin{minipage}{0.80\textwidth}
      
      \def\gAxA{\b\w\A{gAxA}}
      \def\gAxB{\b\w\A{gAxB}}
      \def\gBxA{\op{R_z(-\alpha)}\w\A{gBxA}}
      \def\gBxB{\op{R_z(-\beta)}\w\A{gBxB}}
      \def\gCxA{*-{}\w\A{gCxA}}
      \def\gCxB{*-{}\w\A{gCxB}}
      \def\gDxA{*-{}\w\A{gDxA}}
      \def\gDxB{*-{}\w\A{gDxB}}
      \def\gExA{*-{}\w\A{gExA}}
      \def\gExB{*-{}\w\A{gExB}}
      \def\gFxA{*-{}\w\A{gFxA}}
      \def\gFxB{*-{}\w\A{gFxB}}
      \def\gGxA{*-{}\w\A{gGxA}}
      \def\gGxB{*-{}\w\A{gGxB}}
      \def\gHxA{\op{H}\w\A{gHxA}}
      \def\gHxB{\op{H}\w\A{gHxB}}
      \def\gIxA{\b\w\A{gIxA}}
      \def\gIxB{\b\w\A{gIxB}}
      \def\gJxA{\op{Z}\w\A{gJxA}}
      \def\gJxB{\op{Z}\w\A{gJxB}}
      \def\gKxA{\op{H}\w\A{gKxA}}
      \def\gKxB{\op{H}\w\A{gKxB}}
      \def\gLxA{\meter\w\A{gLxA}}
      \def\gLxB{\meter\w\A{gLxB}}


      \def\bA{\qv{q_{5}}{+}}
      \def\bB{\qv{q_{6}}{+}}


      \xymatrix@R=5pt@C=10pt{
          \bA & \gAxA &\gBxA &\gCxA &\gDxA &\gExA &\gFxA &\gGxA &\gHxA &\gIxA &\gJxA &\gKxA &\gLxA &\N
      \\  \bB & \gAxB &\gBxB &\gCxB &\gDxB &\gExB &\gFxB &\gGxB &\gHxB &\gIxB &\gJxB &\gKxB &\gLxB &\N
      %
      %
      \ar@{-}"gAxA";"gAxB"
      \ar@{-}"gIxA";"gIxB"
      }

      \caption{2-bit Grover Circuit}
      \label{fig:grovercirc}
      \vspace{4mm}
    \end{minipage}
    \begin{minipage}{0.48\textwidth}
      \centering
      \begin{tikzpicture}  
        [scale=.9,auto=center,every node/.style={circle,fill=blue!20}]
          
        \node (a1) at (1,4) {$q_1$};  
        \node (a3) at (3,4) {$q_3$}; 
        \node (a5) at (5,4) {$q_5$}; 
        
        \node (a2) at (1,2) {$q_2$};  
        \node (a4) at (3,2) {$q_4$}; 
        \node (a6) at (5,2) {$q_6$}; 
        
        \draw (a1) -- (a3);
        \draw (a3) -- (a5);  
        
        \draw (a2) -- (a4);
        \draw (a4) -- (a6);  
        
        \draw (a1) -- (a2);
        \draw (a3) -- (a4);  

      \end{tikzpicture}  
      \caption{2-bit Grover Cluster (6-qubit variation) -- corresponds exactly to the circuit in figure \ref{fig:grovercirc}.}
      \label{fig:2bitgrover6}
    \end{minipage}\hfill
    \begin{minipage}{0.48\textwidth}
      \centering
      \begin{tikzpicture}  
        [scale=.9,auto=center,every node/.style={circle,fill=blue!20}]
          
        \node (a1) at (1,4) {$q_1$};  
        \node (a3) at (3,4) {$q_3$};
        
        \node (a2) at (1,2) {$q_2$};  
        \node (a4) at (3,2) {$q_4$};
        
        \draw (a1) -- (a3); 
        
        \draw (a2) -- (a4);
        
        \draw (a1) -- (a2);
        \draw (a3) -- (a4);  

      \end{tikzpicture}  
      \caption{2-bit Grover Cluster (4-qubit variation)}
      \label{fig:2bitgrover4}
    \end{minipage}
    
      \vspace{4mm}
        \begin{lstlisting}
        PrepList([0, 1, 2, 3]);
        CZ(0, 1);
        J(0, 0, 2);
        J(pi, 1, 3);
        CZ(2, 3);
        ReadOut(2, FromAngle(pi));
        ReadOut(3, FromAngle(pi));
        \end{lstlisting}
        \caption{4-qubit variation of Grover's in MCBeth with oracle searching for ``10''}
        \label{fig:grovprog}
  \end{figure}

\section{Simulation and Execution}\label{sec:sim}
In this section, we discuss how we simulate and execute programs in \MCL{} through our own simulator implementation, and how we also compile them down for use on general gate-based quantum systems.
The MCBeth codebase is available online\footnote{\url{https://github.com/seunomonije/mcbeth}} under an open source license.
\subsection{Simulation Implementation}
    We implement \MCL{} using OCaml. We define each individual command as an inductive data type and represent a program as a list of these commands. Importantly, qubits are specified as integers; a constraint on the current implementation is that an initial system must start the qubit number at 0 and then continue numbering successively so that if a system uses $n$ qubits, it will have a qubit associated with each integer between $0$ and $n-1$, inclusively. We use the Lacaml external library\footnote{\url{https://github.com/mmottl/lacaml}} to simulate the execution of a \MCL{} program.


    We provide two simulators for \MCL{} programs: one for weak simulation and one for strong simulation. The key difference between weak and strong simulation is that weak simulation uses randomization to determine which basis to collapse to during measurement and only
    computes that computation path, while strong simulation computes both computation paths and uses the result to construct a probability distribution. Therefore, strong simulation is more computationally intensive because it must compute $2^n$ computation paths when there are $n$ measurements performed, while weak simulation only computes one simulation path. Strong simulation, however, gives a whole picture of the possible outcomes while weak simulation only gives one possible outcome; therefore, in order to extract useful information from a weak simulation, the simulation must be run for many iterations to create a sample of possible outcomes and then construct a probability distribution based on the sample. However, weak simulation is more faithful to actual execution on a quantum computer and is more appropriate for deterministic programs like the Deutsch-Jozsa algorithm.
    
    A weak simulation of an \MCL{} program is executed by passing the program into the \texttt{rand\_eval} function contained in the \texttt{Backend.Run} module; similarly, strong simulation is executed with the \texttt{simulate} function.
    
    These functions return three objects: a matrix representing the output quantum system, an ordered list of qubits corresponding to those qubits in the output quantum system, and a table containing
    entries for the readout qubits. Because the quantum system during the computation gets partially destroyed by the measurement commands, the ordered list of qubits is required in order to properly know the position of each qubit left in the system.

    The return output matrix differs in type for weak and strong simulation. For weak simulation, because we can simply collapse to one basis randomly during the measurement operation, we can perform the simulation using the state vector representation; therefore, a state vector of the output system is returned by \texttt{rand\_eval}. For strong simulation, because we need to combine multiple possibilities of measurement, we perform the simulation using the \textit{density matrix} representation; therefore, a density matrix of the output system is returned by \texttt{simulate}. A density matrix is another way to represent a quantum system mathematically; for unmeasured (\emph{pure}) states, instead of simply having the system represented as a state vector $\ket{\psi}$, we now represent the system as $\ket{\psi}\bra{\psi}$. For probability distributions over measured (\emph{mixed}) states, the corresponding density matrix is the sum of the unscaled outcomes: For instance if we measure $\rho = \ket{+}\bra{+}$ in the computation basis, we get $\ket{0}\bra{0}\rho\ket{0}\bra{0} + \ket{1}\bra{1}\rho\ket{1}\bra{1} = \frac{1}{2}I_2$.
    
\subsection{Removing Measured Qubits}

    Because measurement is a destructive operation, instead of keeping the result as a component in the matrix representing the system, i.e., the  state vector or density matrix, during the measurement operation we can remove it by projecting down into a subspace of a lower dimension. The projection into a smaller subspace removes the qubit from the system while also decreasing the size of a state vector by one-half and the size of a density matrix by three quarters. This allows for future computations to have matrix multiplication on significantly smaller matrices -- allowing for a faster simulation of the quantum circuit. For example, in the case of our teleportation example in Figure \ref{fig:teleprog}, each measurement command will remove a qubit from the system; therefore, we no longer need to keep track of it in our matrix.

  \subsection{Serialization and cross compatibility}
  In order to enable exploration of programs written in \MCL{} to other formats,
  \MCL{} serializes down to a standard JSON format which can be parsed and
  translated to other frameworks as needed.
  
  Serialization of the language is straightforward. Each \MCL{} command maps
  to a JSON object with the name of the command as the key, and a nested 
  JSON object as a value. This nested JSON object contains relevant
  information about the command following the structure of the \textit{cmd}
  type in our language implementation. For example, the \MCL{} command:
 
  \centerline{\textit{Entangle(0, 1)}}
  
  translates to the JSON object:
  
 \centerline{ \textit{\{'Entangle': \{'on\_qubits': [0, 1]\}\}}}
 
 As programs get larger, we represent them with a list of these objects, which
 can be parsed and converted into the desired language. In our work, we
 provide a translator from \MCL{} to Cirq with the \textit{CirqBuilder} class. 
 With this translation implementation, we can translate \MCL{} programs 
 into OpenQASM by running them through Cirq first. The compiled
 OpenQASM representation of \MCL{} programs makes the language compatible with
 in almost every major quantum programming framework
 or language, and also allows us to run programs written in 
 \MCL{} on real quantum hardware.
 The programs that execute on the quantum devices will not be MCBeth programs  -- as measurement-based quantum computers aren't readily available -- but they will correspond semantically to the original program.

  \subsection{Dependent measurements in translation}
   
 The caveat to converting programs following a measurement based
model to be compatible with gate based frameworks is handling the concept of 
\textit{dependent measurement}. Dependent measurements are a fundamental 
component of quantum computation via the measurement calculus; however,
in gate-based models, true dependent measurement is impossible without storing measurement results of one qubit in a classical register, and then using those results to perform an operation on another qubit. This is a limitation both in programming languages, and in hardware. This presents a problem with translation, since \MCL{} programs should natively support dependent measurement and the languages we translate to do
not.


To solve this problem, we leverage the \textit{deferred measurement principle} to convert
all instances of dependent measurements in an \MCL{} program to an equivalent
representation compatible with gate based models. The deferred measurement 
principle allows measurements to be moved around to convenient locations in a circuit 
with no effect on the probability distribution of the measurement outcomes. This essentially means that programs translated from \MCL{} to OpenQASM do not include true dependent measurement, however, the results of running programs will be equivalent.



\section{MCBeth-specific Architectures}\label{sec:arch}
    In this section, we outline some systems and computer architectures that can take advantage of measurement-based computation's unique benefits, and also how our language \MCL{} can be used to realize and work with these systems.
    
     \subsection{Distributed Computation}\label{sec:distributed}
      
      MBQC combined with quantum networking to perform distributed computations produces the greatest potential application of \MCL{}. Because \MCL{} standardizes the program to have entanglement first, \MCL{} programs could be initialized on a central ``entangling'' node and, then, via quantum channels (for instance, the photonic channels proposed by Monroe et al.~\cite{monroe2014}), qubits could be sent to different nodes. Because the qubits are entangled, operations on one entangled qubit will still affect the qubits with which it is entangled, regardless of where these qubits are. Moreover, because the signals are classical pieces of information, they may be shared accross classical channels during the computation. This would then allow the measurement and corrections sections of the computation to be done distributedly, removing the need for each node to have the ability to store and operate on large numbers of qubits. While this could be done in the gate-based model by converting non-suitable circuits into a distributed form, programs in \MCL{} are created by default in a distributed form; therefore, there is no need to convert the program; This makes \MCL{} more natural to program distributed programs in. 
      
      Notably, each edge node could be specialized in performing either entanglement, measurement, or corrections, allowing for hardware resources to be specialized to the operations the node supports. The central entangling node would be made specifically for the purpose of entanglement and, therefore, would not need to support measurement or correction operations. Similarly, the nodes for measurement and corrections would only need to handle their respective operations and, therefore, their hardware could be designed to optimize performance in their respective domains.
    
      One of the hardest challenges in achieving scalable quantum computation is the issue of having quantum computers handle large numbers of qubits. By using \MCL{} to create programs that can be distributed to multiple quantum computers, the issue of scaling the number of qubits a computer can handled is isolated to the central entangling node. Specifically, since the entangling node is the only one which requires all qubits to be on the same computer in order to perform the operation, this is the only node which needs to handle a large number of qubits. Furthermore, because the entangling node does not handle measurement or corrections, this distributed approach, unlike today's quantum computers, will only need to scale in regards to performing entanglement and distribution of the qubits.
    
      Not only does distributed computation allow for the decrease in qubits per edge node but it also allows for the execution of certain steps of a program in parallel. Figure~\ref{fig:optimalsub} displays just this case; there are two subsystems: $S_1 = [q_1, q_2, q_3, q_4]$ and $S_2 = [q_5, q_6, q_7, q_8]$. Because no qubit in $S_1$ is dependent on a qubit in $S_2$, and vice versa, the computation of each subsystem can be done perfectly in parallel.
    
  \begin{figure}
    \begin{minipage}{0.48\textwidth}
      \begin{tikzpicture}  
        [scale=.9,auto=center,every node/.style={circle,fill=blue!20}]
          
        \node (a1) at (1,2) {$q_5$};  
        \node (a2) at (2.8,2) {$q_6$}; 
        \node (a3) at (4.6,2) {$q_7$};  
        \node (a4) at (6.4,2) {$q_8$};  
        \node (a5) at (1,4) {$q_1$};
        \node (a6) at (2.8,4) {$q_2$}; 
        \node (a7) at (4.6,4) {$q_3$}; 
        \node (a8) at (6.4,4) {$q_4$}; 
        
        \draw (a1) -- (a2);
        \draw (a2) -- (a3);  
        \draw (a3) -- (a4);  
        \draw (a5) -- (a6);  
        \draw (a6) -- (a7); 
        \draw (a7) -- (a8);  
        
        \draw[dashed] (3.8, 2) ellipse (110pt and 22pt);
        \draw[dashed] (3.8, 4) ellipse (110pt and 22pt);

      \end{tikzpicture}  
      \caption{Optimal Subsystem Partitioning}
      \label{fig:optimalsub}
    \end{minipage}\hfill
    \begin{minipage}{0.48\textwidth}
      \begin{tikzpicture}  
        [scale=.9,auto=center,every node/.style={circle,fill=blue!20}]
          
        \node (a1) at (1,2) {$q_5$};
        \node (a2) at (3,2) {$q_6$}; 
        \node (a3) at (5,2) {$q_7$};  
        \node (a4) at (7,2) {$q_8$};  
        \node (a5) at (1,4) {$q_1$};
        \node (a6) at (3,4) {$q_2$}; 
        \node (a7) at (5,4) {$q_3$}; 
        \node (a8) at (7,4) {$q_4$}; 
        
        \draw (a1) -- (a2);
        \draw (a2) -- (a3);  
        \draw (a3) -- (a4);  
        \draw (a5) -- (a6);  
        \draw (a6) -- (a7); 
        \draw (a7) -- (a8);
        
        \draw[dashed] (1, 3) ellipse (22pt and 60pt);
        \draw[dashed] (3, 3) ellipse (22pt and 60pt);
        \draw[dashed] (5, 3) ellipse (22pt and 60pt);
        \draw[dashed] (7, 3) ellipse (22pt and 60pt);
  
      \end{tikzpicture}  
      \caption{Non-Optimal Subsystem Partitioning}
      \label{fig:nonoptimalsub}
    \end{minipage}
  \end{figure}

      We provide a module \texttt{Backend.Distributed} to split a \MCL{} program up into distributed subsystems; this is done by passing a program into \texttt{build\_dist\_prog} along with a list of lists containing how the qubits should be partitioned. This module contains utility functions to split a program into isolated subprograms for each node in a network. It is up to the designers of the architecture to implement the communication between nodes.
      
     \subsection{\MCL{} for MBQC-Specific Architecture}
  Using \MCL{} to write programs for MBQC-specific architectures would provide a high-level programming abstraction with low-level control, assuming a mapping between \MCL{} commands and the corresponding pulses and wavelengths is created for the specific hardware.
  
  In gate-based models, this mapping between the commands and the hardware is achieved by compilation, which maps input circuits to the topology of the hardware, and \textit{pulse scheduling}, which configures where and when pulses are sent to the individual qubits. With a variety of gates and operations, compiling and scheduling gate-based circuits can become a complex and involved process. In measurement-based models and specifically cluster-state computing, compiling the circuits down is straightforward.
  
  The separation of entanglement and measurement in cluster-state, one-way computing models makes it easier for architectures to delegate hardware specifically for entangling qubits and creating the cluster-state, then send pulse schedules to that cluster without needing to worry about re-entangling or any other operations that are not measurement operations. An example of this architecture can be seen in a 2014 experimental implementation of Simon's algorithm \cite{simons-experiment}. The group was able to modify an existing photonic architecture used for a quantum error correction code to generate a linear 1D cluster state by producing photon pairs and entangling them using a polarizing beam splitter (PBS). They then sent pulses to these entangled photons to measure in the X, Y, or Z basis, with distinct pulse paths for each basis. 

  In theory, all quantum algorithms that could be represented through the cluster state, one-way model could be run on this architecture, assuming that the hardware could produce and entangle enough photons for an $n$ qubit cluster state. Compiling all quantum algorithms in this model becomes simple as well: prepare the cluster state, and schedule each measurement one after another. We could also use the compiler to optimize inputs which would not be needed as any changes in the algorithm, which would be represented in \MCL{}, would rely only on changes in measurement. This allows the programmer to be able to develop algorithms at a high-level while also understanding the program at the pulse level. In many cases, pulse-level understanding would not be necessary, however; this shows that from the programmer's point of view, understanding the computation at all levels in \MCL{} is significantly easier than doing so in the gate-based model.
  
  Additionally, MQBC architectures such as the one in the experiment mentioned above provide versatility when implementing algorithms while maintaining the same hardware-level control. All algorithms outlined in Section 8.1 would be compatible with the architecture provided in the paper, and new algorithms could be easily explored at both the algorithmic and pulse level without any modifications to the existing hardware. 
  



 
\section{Evaluation}\label{sec:eval}

In the evaluation of our language, we answer the questions:

\begin{enumerate}
    \item What can we program in MCBeth that is difficult to program via gate-based languages?
    \item What are the advantages of measurement-based quantum algorithms versus gate-based, and how does our language enable programmers to use them?
    \item How can measurement-based programs make quantum programming easier in both small and large codebases?
    \item How well does our language integrate with existing quantum programming languages?
    \item How efficiently can we simulate distributed \MCL{} programs?
\end{enumerate}

To evaluate \MCL{}, we conducted a literature search and implemented several measurement-based quantum programs from our findings using \MCL{}. We then implemented the same algorithms using IBM's gate-based language Qiskit \cite{ibmq}, and documented the difference in qubit and operation count between the measurement and gate based versions of the programs. We also analyzed code snippets of one of the most fundamental quantum protocols, quantum teleportation, in \MCL{} and in Qiskit, in order to assess the implementation benefits of \MCL{} as a language and measurement-based computation as a concept. Finally, we compiled down a handful of quantum programs to the gate based model and ran them on IBM's \textit{ibmq-belem} machine, a 5-qubit quantum processor with a quantum volume of 16, supporting 2500 circuit layer operations per second. 

Through our evaluation, we discovered:
\begin{itemize}
    \item Writing programs which require dependent measurements using \MCL{} is significantly easier than gate-based counterparts due to \MCL{}'s native support for signals.
    \item Quantum programs written in  \MCL{} are simpler and require less code.
    \item Overall programming overhead is less when developing algorithms in \MCL{} due to the smaller set of operations, allowing programmers to focus on developing quantum algorithms and worry less about bugs.
    \item \MCL{} can successfully compile down to a gate-based representation and can be run on a public quantum system.
\end{itemize}

  \subsection{Implementation of Algorithms}
To show the usefulness of our language, we conducted a literature search to find various measurement-based quantum algorithms and programs and implemented them using \MCL{}. We then implemented the equivalent gate-based algorithms using IBM's Qiskit framework. In the table below, we provide information comparing the qubit count and operation count between the equivalent gate-based and measurement-based algorithms.\footnote{
      Gate-based algorithms were implemented in Qiskit, using Qiskit Textbook as a reference \cite{Qiskit-textbook}. 
      Our \MCL{} programs used existing measurement-based descriptions of
      Deutsch-Jozsa \cite{bv-dj-tame},
      Simon's Algorithm \cite{simons-experiment},
      Bernstein-Vazirani \cite{bv-dj-tame},
      QEC Phase Flip \cite{trapped-ions}, and the
      QFT \cite{Raussendorf_2003},
      }


\newcommand{\ra}[1]{\renewcommand{\arraystretch}{#1}} 
     \begin{figure}[h]\centering
     \ra{1.2}
    \begin{tabular}{@{}cccccc@{}}\toprule
      \shortstack{Algorithm/Program} & 
     \shortstack{Qubits\\(Qiskit)} & 
     \shortstack{Gate count\\(Qiskit)} & 
     \shortstack{Qubits\\(\MCL{})} & 
    \shortstack{Ops\\(Cluster)} & 
    \shortstack{Ops\\(Algorithm)} \\  \midrule
     Deutsch-Jozsa ($n$-bit)  & $n+1$ & $2n + 1 + O$ & $2 + 2n$ & $4 + 4n$ & $n+1$\\
     
     Simon's ($n$-bit)  & $2n$ & $2n + O$ & $n^2 - n + 1$  & $2n^2 - 2n + 2$ & $n$\\
     
     Bernstein-Vazirani ($n$-bit)& $n + 1$ & $2n + O$ & $2 + 2n$ & $4 + 4n$ & $n+1$\\ 
     
     
     QEC Phase Flip ($n$-bit) & $n+1$ & $4n - 1 + O$ & $n+2$ & $2n$ & $n+2$ \\
     
     QFT ($n$-bit)  & $n$ & $2n^2$ & $n$ & $2n^2$ & $n$\\ 
     
     
     
    Grover's (2-bit) & 2 & 10 & 6 & 10 & 2\\
     Grover's (2-bit alt) & 2 & 10 & 4 & 7 & 2 \\ 
     
     Teleportation (Example) & 3 & 8 & 3 & 2 & 4\\ 
\bottomrule
    \end{tabular}
        \caption{Table of algorithms implemented in \MCL{}, including qubit and operation count for gate and measurement based variations of an algorithm.}
         \label{tab:my_label}
    \end{figure}
    
     Note that the gate and operation counts do not include the gates of decomposed oracles. We represent oracles with $O$. We split measurement-based operation counts into two categories, cluster state generation, and algorithm measurements, written as Ops (Cluster) and Ops (Algorithm) respectively. For operation counts, note that we represent the maximum possible for that program.

  \subsection{Comparing \MCL{} code to Qiskit code}
  
    We present implementations of the quantum teleportation protocol in \MCL{} and Qiskit in Section \ref{app:codecmp} of the Appendix, highlighting significant differences in available features and amount of code when programming one of the most fundamental protocols in quantum computing. The Qiskit implementations are based off of work from IBM's Qiskit Textbook.

  Comparing these two programs highlights significant differences in both programming with \MCL{} vs. Qiskit, and in gate-based and measurement-based computation. 
  

  The first difference is obvious: programming measurement-based algorithms is easier to do in \MCL{} then it is in Qiskit, or any other gate-based languages. As shown in the prior sections, \MCL{} can seamlessly implement cluster states and measurement-based programs.

  Elaborating on this point, programmers working in \MCL{} only need to concern themselves with 
  measuring qubits, choosing the order, angle, and signals for each measured qubit.
  Setting up the cluster state requires knowledge of only two operations ($CZ$ or $J$), and so developing powerful quantum algorithms in \MCL{} can be reduced to these two steps. Compared to the gate-based model, where programmers need to keep track of 
  significantly more gates and the interleaving single-qubit gates, entangling and measurements,
  programming in \MCL{} holds a significant simplicity advantage.
  
  Another advantage when building in \MCL{} and measurement-based languages is that ancilla qubits a non-factor; the majority of the time they are built into the cluster and are taken care of within the algorithm. The few ancilla (or scratch) qubits that appear in a \MCL{} program are allocated at the beginning of the program and do not have any influence on the algorithm. When writing an algorithm in the measurement-based model, it's common for clusters to be prepared with designated readout qubits, which may change depending on what the user wants to get out of the algorithm. This is shown in our examples of measurement-based algorithms in Section \ref{sec:programmingMCB}. Ancilla qubits are widely used in gate-based algorithms and protocols, and not only are they an extra step to manage when writing quantum algorithms, but ancilla deallocation errors may propagate throughout the program, so avoiding them can make a programmer's life easier.

  Finally, the difference in codebase sizes needed to implement programs in the measurement-based and gate-based models is something to consider. Using \MCL{} and measurement-based algorithms to write and test quantum algorithms can lead to a considerably smaller and simpler codebases at scale due the measurement based model's finite operation set and \MCL{}'s native ability to handle more involved concepts like cluster state preparation and dependent measurements. 
  

  \subsection{Running \MCL{} programs on a real quantum system}

  To demonstrate an end-to-end use of our language, we present
  results from running 4 programs from our language on a real
  quantum system via the IBM Quantum Experience \cite{ibmq}. 
  Results are from running translated \MCL{} circuits on the ibmq\_belem
  machine. We emphasize that this is a compilation of the following programs from \MCL{} to a gate-based representation, as mentioned in Section 5.2. All times are measured in seconds and plotted on Figure \ref{fig:runtimes}.

\begin{figure}[h]
\centering
\begin{minipage}{0.48\textwidth}

    \scalebox{0.7}{%
    \begin{tikzpicture}
      \begin{axis} [xbar = .05cm,
        bar width = 12pt,
        xmin = 0, 
        xmax = 16, 
        ytick={1, 2, 3, 4},
        yticklabels={1-qubit\\ teleportation, X correction, Z correction, 4-qubit 2-bit \\Grover's\\ search},
        yticklabel style={align=right},
        legend style={at={(0.7,1.2)},anchor=north},
        nodes near coords,
        enlarge y limits = {abs = .8}
    ]
    
    \addplot coordinates {(11.3,1) (6.4,2) (10.8,3) (13.7,4)};
    \addplot coordinates {(4.0,1) (4.1,2) (4.4,3) (4.7,4)};
    
    \legend{Total system runtime (s), Total time in system (s)}
    \end{axis}
\end{tikzpicture}}
\caption{Runtimes from compiled \MCL{} cirucuits on real systems}
\label{fig:runtimes}
  
\end{minipage}
\begin{minipage}{0.04\textwidth}
\end{minipage}
\begin{minipage}{0.48\textwidth}

  \begin{figure}[H]
    \centering
    \scalebox{0.7}{%
    \begin{tikzpicture}
      \begin{axis}[
        title={Runtime of Distributed Cluster Simulation},
        xlabel={Number of qubits per subsystem},
        ylabel={Number of subsystems},
        zlabel={Execution time in seconds},
        x dir=reverse,
      ]
      \addplot3[
          surf,
      ] 
      coordinates {
        (1, 1, 0.000396)
        (2, 1, 0.000182)
        (3, 1, 0.000186)
        (4, 1, 0.000163)
        (5, 1, 0.000151)
        (6, 1, 0.000116)
        (7, 1, 0.012198)
        (8, 1, 0.039812)
        (9, 1, 0.174244)
        (10, 1, 0.902514)

        (1, 2, 0.000342)
        (2, 2, 0.000694)
        (3, 2, 0.000324)
        (4, 2, 0.000397)
        (5, 2, 0.001452)
        (6, 2, 0.007304)
        (7, 2, 0.024862)
        (8, 2, 0.06954)
        (9, 2, 0.335245)
        (10, 2, 1.685161)

        (1, 3, 0.000952)
        (2, 3, 0.001495)
        (3, 3, 0.001631)
        (4, 3, 0.001676)
        (5, 3, 0.010795)
        (6, 3, 0.018736)
        (7, 3, 0.021325)
        (8, 3, 0.121542)
        (9, 3, 0.597372)
        (10, 3, 2.690226)

        (1, 4, 0.000852)
        (2, 4, 0.004372)
        (3, 4, 0.002776)
        (4, 4, 0.006248)
        (5, 4, 0.014803)
        (6, 4, 0.028682)
        (7, 4, 0.067125)
        (8, 4, 0.18675)
        (9, 4, 0.793009)
        (10, 4, 3.637459)

        (1, 5, 0.00124)
        (2, 5, 0.002796)
        (3, 5, 0.004844)
        (4, 5, 0.006785)
        (5, 5, 0.019122)
        (6, 5, 0.032683)
        (7, 5, 0.080468)
        (8, 5, 0.235182)
        (9, 5, 1.002712)
        (10, 5, 4.839309)

        (1, 6, 0.001379)
        (2, 6, 0.003309)
        (3, 6, 0.004948)
        (4, 6, 0.01301)
        (5, 6, 0.019803)
        (6, 6, 0.044572)
        (7, 6, 0.105332)
        (8, 6, 0.273205)
        (9, 6, 1.231518)
        (10, 6, 5.830352)

        (1, 7, 0.001478)
        (2, 7, 0.003312)
        (3, 7, 0.00423)
        (4, 7, 0.017226)
        (5, 7, 0.028057)
        (6, 7, 0.051243)
        (7, 7, 0.138705)
        (8, 7, 0.34399)
        (9, 7, 1.491673)
        (10, 7, 6.848694)

        (1, 8, 0.001811)
        (2, 8, 0.006213)
        (3, 8, 0.012076)
        (4, 8, 0.016111)
        (5, 8, 0.028442)
        (6, 8, 0.056722)
        (7, 8, 0.161498)
        (8, 8, 0.46985)
        (9, 8, 1.886204)
        (10, 8, 7.76505)

        (1, 9, 0.00207)
        (2, 9, 0.010973)
        (3, 9, 0.012822)
        (4, 9, 0.014285)
        (5, 9, 0.034208)
        (6, 9, 0.059563)
        (7, 9, 0.160437)
        (8, 9, 0.428775)
        (9, 9, 2.205881)
        (10, 9, 8.815954)

        (1, 10, 0.001865)
        (2, 10, 0.010617)
        (3, 10, 0.010951)
        (4, 10, 0.020509)
        (5, 10, 0.04917)
        (6, 10, 0.069592)
        (7, 10, 0.164654)
        (8, 10, 0.631766)
        (9, 10, 2.251892)
        (10, 10, 10.305388)

      };
      \end{axis}
    \end{tikzpicture}
    }
    \caption{Distributed Simulation}
    \label{fig:dist}
  \end{figure} 
     
\end{minipage}

\end{figure}
  
\subsection{Distributed \MCL{} simulation}

      While we cannot simulate accurately all distributed computations because entanglement allows entangled qubits split between different nodes to affect each other, if we split a computation up in such a way where each node in the network has only qubits entangled with other qubits in the same node, entanglement does not exist between nodes; therefore, an accurate distributed simulation can be performed in such cases. To measure simulation performance, we ran weak simulations of arbitrary programs utilizing $n$ qubits per node with $m$ nodes. Each node, therefore, contained one subsystem of the computation. Furthermore, each node ran a computation of the form akin to teleportation: a linear cluster. Therefore, each computation as a whole will produce $m$ random output qubits. This grouping is like that shown in Figure~\ref{fig:optimalsub} with each subsystem circled. Results from the classical weak simulation of distributed programs of various configurations on a single processor is shown in Figure~\ref{fig:dist} -- each distributed node was simulated in a separate thread.
    
      Say that a computation on $n$ qubits in one subsystem takes $t_n$ time. Since the simulation will run in parallel with $m$ subsystems, the execution time of the simulation will take at least $m \cdot t_n$ time to complete when run sequentially on the same processor. In Figure \ref{fig:dist}, this linear increase in computation time is best seen in the subsystems which contain ten qubits; as the number of subsystems increase, we can see a clear linear increase in computation time. In the non-distributed equivalent program, this would take an exponential amount of time instead. If we were to run each cluster concurrently on separate processors, then the total time would simply be $t_n$.
    
      Importantly, these simulations present the best case scenario: each subsystem does not have commands dependent on qubits in other subsystems; therefore, each computation can be performed perfectly in parallel, not having to wait for measurement information in other subsystems. If, for example, we regroup the qubits into subsystems as shown in Figure~\ref{fig:nonoptimalsub}, the computations of each subsystem $S_i$ would have to wait for the preceding subsystem $S_{i-1}$ to finish because $S_i$ has commands dependent on qubits in $S_{i-1}$; also, note that in this grouping, we decrease the number of qubits per subsystem but also introduce entangling between subsystems, which will lead to an incorrect classical distributed simulation.

\section{Related Work}\label{sec:relatedwork}

\paragraph{Algorithmic Languages} One attempt to move beyond the gate-based model includes algorithmic languages, such as Q\# \cite{q-sharp} and IQu \cite{iqu}, which aren't tied to any specific model.  
They move beyond specific quantum models and aim to reflect our standard, higher-level programming languages. These languages, however, still must be compiled eventually down to some lower-level quantum language. Some cannot be compiled at all yet, and others like Q\# compile to gate-based languages. The caveat with these languages, however, is that they do not teach the programmer how to construct a quantum program since they attempt to make quantum programming work the same as classical programming. These languages don't give programmers a good model for realizable quantum hardware, which potentially limits their utility no intuition for the core of quantum computation is built. 

As the quantum computing field grows, these types of programming languages may become more useful. However, in present day, building, designing, and running meaningful algorithms on quantum computers require a working knowledge of the underlying hardware, especially because there are so many different ways these quantum systems can be built. An algorithm written in a language may only be compatible on a certain type of system, so working in languages with hardware awareness will be important in the forseeable future. This is a benefit of \MCL{} as compared to these other languages; with \MCL{}, the programmer gets higher-level simplicity with hardware-level awareness; in contrast, with more algorithmic languages, programmer gets higher-level simplicity at the expense of hardware-level awareness. 

\paragraph{ZX-calculus} Another alternative to the gate-based model is the ZX-calculus\cite{vandewetering2020zxcalculus}, a graphical representation of programs which aims to challenge the conventional circuit diagram. The ZX-calculus distinguishes itself from other methods of quantum computation by introducing both representations and computations that aren't possible or apparent in the gate-based representation. Much like the measurement calculus, the ZX-calculus simplifies instructions to a limited set of commands and offers a set of rewriting rules to simplify large and complex circuits both conceptually and in implementation.

Since the ZX-calculus originated from early work in MBQC, measurement-based programs can and have been represented using the model. However, these ZX programs were created from previously existing programs either in gate or measurement-based representations and were not originally designed using the ZX-calculus. 
This is a direct contrast from our language; we aim to provide a method of representing, designing, and running measurement based programs utilizing the same fundamental structure as prior work on MBQC and quantum computing in general, with a few additional tools to make programming and conceptualizing programs easier. 
This contrast is important to note. While the ZX calculus is a powerful and extremely useful tool to reason and work with MBQC programs, 
it is not intended to perform the role of a programming language.

\paragraph{Paddle Quantum} Another work on MBQC languages released simultaneously to ours is Paddle Quantum, a python framework, which offers a sub-module that supports measurement-based quantum computing \cite{paddle}. This work contains both similarities and differences to the work done in this paper. The similarities are straightforward, using Paddle Quantum's MBQC package, one can prepare cluster states and simulate them, offering users the ability to use measurement calculus commands. The differences are more specific. Paddle Quantum offers an automatic compiler from gate-based programs to measurement-based programs, but does not do the reverse; \MCL{} on the other hand does allow for measurement-based to gate-based translation but not the reverse. Since the MBQC functionality of Paddle Quantum is a sub-module of a larger Python package, and not a standalone package like \MCL{}, it is more verbose than \MCL{} as it provides more complex naming conventions for the different operations. Paddle Quantum does, however, provide a more flexible naming convention for qubits during program construction. Finally, Paddle Quantum's MBQC module also performs additional optimizations for weak simulation; Paddle Quantum also only support weak simulation while \MCL{} supports both weak and strong simulation.

\section{Conclusion}
Most importantly, we've argued that programming in MCBeth is more intuitive than doing so in the gate-based model because MCBeth's commands are tied closer to the actual quantum operations being performed. Prior to this work, measurement-based quantum computing was known to be a powerful alternative model to gate-based quantum computation, but it has rarely been used for practical quantum programming. In developing \MCL{}, we were able to explore the advantages and disadvantages of this model from a programmer's perspective. We provided several examples of measurement-based programs written in \MCL{} and compared them to their gate-based counterparts. We created an open source repository containing a compiler and simulator for the easy creation and simulation of MCBeth programs; we also provide a compiler from \MCL{} to a gate-based model, allowing \MCL{} programs to be run on common, real quantum hardware, or imported into another gate-based quantum programming language. We also argued that programming in MCBeth allows one to naturally create distributed quantum algorithms, which could be executed on a network of machines where each machine is specialized to handle a different type of quantum operation.

For future work, we hope to extend MCBeth by introducing a generalized J operator to the measurement calculus which would allow multiple qubits to be entangled and corrected based on one measured qubit; and extend MCBeth to handle the \textit{extended measurement calculus} \cite{extended-meas-calc}, which would increase the range of states the measurement operator could measure in and, thus, allow for more flexibility in creating MCBeth programs.
Additionally, we also hope to create a tool for the automatic creation of MCBeth programs via the diagrammatic construction of cluster states. 
Future directions could also include the execution of MCBeth programs directly on existing MBQC hardware and on future hardware specialized for distributed MBQC, as well as further integrations with existing toolchains such as Qiskit and ZX. 





\label{mylastpage}

\begin{acks}                            
  This work is funded in part by \href{https://www.epiqc.cs.uchicago.edu/}{EPiQC}, an NSF Expedition in Computing, under Award Number \href{https://www.nsf.gov/awardsearch/showAward?AWD_ID=1730449}{CCF-1730449}.
\end{acks}

\bibliography{main}


\begin{thebibliography}{22}


\ifx \showCODEN    \undefined \def \showCODEN     #1{\unskip}     \fi
\ifx \showDOI      \undefined \def \showDOI       #1{#1}\fi
\ifx \showISBNx    \undefined \def \showISBNx     #1{\unskip}     \fi
\ifx \showISBNxiii \undefined \def \showISBNxiii  #1{\unskip}     \fi
\ifx \showISSN     \undefined \def \showISSN      #1{\unskip}     \fi
\ifx \showLCCN     \undefined \def \showLCCN      #1{\unskip}     \fi
\ifx \shownote     \undefined \def \shownote      #1{#1}          \fi
\ifx \showarticletitle \undefined \def \showarticletitle #1{#1}   \fi
\ifx \showURL      \undefined \def \showURL       {\relax}        \fi
\providecommand\bibfield[2]{#2}
\providecommand\bibinfo[2]{#2}
\providecommand\natexlab[1]{#1}
\providecommand\showeprint[2][]{arXiv:#2}

\bibitem[\protect\citeauthoryear{??}{pad}{2020}]%
        {paddle}
 \bibinfo{year}{2020}\natexlab{}.
\newblock \bibinfo{title}{{Paddle Quantum}}.
\newblock
\newblock
\urldef\tempurl%
\url{https://github.com/PaddlePaddle/Quantum}
\showURL{%
\tempurl}


\bibitem[\protect\citeauthoryear{Abbas, Andersson, Asfaw, Corcoles, Bello,
  Ben-Haim, Bozzo-Rey, Bravyi, Bronn, Capelluto, Vazquez, Ceroni, Chen, Frisch,
  Gambetta, Garion, Gil, Gonzalez, Harkins, Imamichi, Jayasinha, Kang,
  h.~Karamlou, Loredo, McKay, Maldonado, Macaluso, Mezzacapo, Minev, Movassagh,
  Nannicini, Nation, Phan, Pistoia, Rattew, Schaefer, Shabani, Smolin, Stenger,
  Temme, Tod, Wanzambi, Wood, and Wootton.}{Abbas et~al\mbox{.}}{2020}]%
        {Qiskit-textbook}
\bibfield{author}{\bibinfo{person}{Amira Abbas}, \bibinfo{person}{Stina
  Andersson}, \bibinfo{person}{Abraham Asfaw}, \bibinfo{person}{Antonio
  Corcoles}, \bibinfo{person}{Luciano Bello}, \bibinfo{person}{Yael Ben-Haim},
  \bibinfo{person}{Mehdi Bozzo-Rey}, \bibinfo{person}{Sergey Bravyi},
  \bibinfo{person}{Nicholas Bronn}, \bibinfo{person}{Lauren Capelluto},
  \bibinfo{person}{Almudena~Carrera Vazquez}, \bibinfo{person}{Jack Ceroni},
  \bibinfo{person}{Richard Chen}, \bibinfo{person}{Albert Frisch},
  \bibinfo{person}{Jay Gambetta}, \bibinfo{person}{Shelly Garion},
  \bibinfo{person}{Leron Gil}, \bibinfo{person}{Salvador De La~Puente
  Gonzalez}, \bibinfo{person}{Francis Harkins}, \bibinfo{person}{Takashi
  Imamichi}, \bibinfo{person}{Pavan Jayasinha}, \bibinfo{person}{Hwajung Kang},
  \bibinfo{person}{Amir h. Karamlou}, \bibinfo{person}{Robert Loredo},
  \bibinfo{person}{David McKay}, \bibinfo{person}{Alberto Maldonado},
  \bibinfo{person}{Antonio Macaluso}, \bibinfo{person}{Antonio Mezzacapo},
  \bibinfo{person}{Zlatko Minev}, \bibinfo{person}{Ramis Movassagh},
  \bibinfo{person}{Giacomo Nannicini}, \bibinfo{person}{Paul Nation},
  \bibinfo{person}{Anna Phan}, \bibinfo{person}{Marco Pistoia},
  \bibinfo{person}{Arthur Rattew}, \bibinfo{person}{Joachim Schaefer},
  \bibinfo{person}{Javad Shabani}, \bibinfo{person}{John Smolin},
  \bibinfo{person}{John Stenger}, \bibinfo{person}{Kristan Temme},
  \bibinfo{person}{Madeleine Tod}, \bibinfo{person}{Ellinor Wanzambi},
  \bibinfo{person}{Stephen Wood}, {and} \bibinfo{person}{James Wootton.}}
  \bibinfo{year}{2020}\natexlab{}.
\newblock \bibinfo{title}{Learn Quantum Computation Using Qiskit}.
\newblock
\newblock
\urldef\tempurl%
\url{http://community.qiskit.org/textbook}
\showURL{%
\tempurl}


\bibitem[\protect\citeauthoryear{{Cirq Developers}}{{Cirq Developers}}{2018}]%
        {cirq}
\bibfield{author}{\bibinfo{person}{{Cirq Developers}}.}
  \bibinfo{year}{2018}\natexlab{}.
\newblock \bibinfo{title}{Cirq}.
\newblock
\newblock
\urldef\tempurl%
\url{https://doi.org/10.5281/zenodo.4062499}
\showDOI{\tempurl}


\bibitem[\protect\citeauthoryear{Cross, Bishop, Smolin, and Gambetta}{Cross
  et~al\mbox{.}}{2017}]%
        {cross2017open}
\bibfield{author}{\bibinfo{person}{Andrew~W. Cross}, \bibinfo{person}{Lev~S.
  Bishop}, \bibinfo{person}{John~A. Smolin}, {and} \bibinfo{person}{Jay~M.
  Gambetta}.} \bibinfo{year}{2017}\natexlab{}.
\newblock \bibinfo{title}{Open Quantum Assembly Language}.
\newblock
\newblock
\showeprint[arxiv]{1707.03429}~[quant-ph]


\bibitem[\protect\citeauthoryear{Danos, Kashefi, and Panangaden}{Danos
  et~al\mbox{.}}{2007}]%
        {danos}
\bibfield{author}{\bibinfo{person}{Vincent Danos}, \bibinfo{person}{Elham
  Kashefi}, {and} \bibinfo{person}{Prakash Panangaden}.}
  \bibinfo{year}{2007}\natexlab{}.
\newblock \showarticletitle{The Measurement Calculus}.
\newblock \bibinfo{journal}{\emph{J. ACM}} \bibinfo{volume}{54},
  \bibinfo{number}{2} (\bibinfo{date}{apr} \bibinfo{year}{2007}),
  \bibinfo{pages}{8–es}.
\newblock
\showISSN{0004-5411}
\urldef\tempurl%
\url{https://doi.org/10.1145/1219092.1219096}
\showDOI{\tempurl}


\bibitem[\protect\citeauthoryear{Danos, Kashefi, Panangaden, and Perdrix}{Danos
  et~al\mbox{.}}{2009}]%
        {extended-meas-calc}
\bibfield{author}{\bibinfo{person}{Vincent Danos}, \bibinfo{person}{Elham
  Kashefi}, \bibinfo{person}{Prakash Panangaden}, {and} \bibinfo{person}{Simon
  Perdrix}.} \bibinfo{year}{2009}\natexlab{}.
\newblock \bibinfo{booktitle}{\emph{Extended Measurement Calculus}}.
\newblock \bibinfo{publisher}{Cambridge University Press},
  \bibinfo{address}{United States}, \bibinfo{pages}{235--310}.
\newblock
\urldef\tempurl%
\url{https://doi.org/10.1017/CBO9781139193313.008}
\showDOI{\tempurl}


\bibitem[\protect\citeauthoryear{Deutsch}{Deutsch}{1985}]%
        {deutsch}
\bibfield{author}{\bibinfo{person}{David Deutsch}.}
  \bibinfo{year}{1985}\natexlab{}.
\newblock \showarticletitle{Quantum theory, the Church\&\#x2013;Turing
  principle and the universal quantum computer}.
\newblock \bibinfo{journal}{\emph{Proceedings of the Royal Society of London.
  A. Mathematical and Physical Sciences}} \bibinfo{volume}{400},
  \bibinfo{number}{1818} (\bibinfo{year}{1985}), \bibinfo{pages}{97--117}.
\newblock
\urldef\tempurl%
\url{https://doi.org/10.1098/rspa.1985.0070}
\showDOI{\tempurl}
\showeprint{https://royalsocietypublishing.org/doi/pdf/10.1098/rspa.1985.0070}


\bibitem[\protect\citeauthoryear{Deutsch and Jozsa}{Deutsch and Jozsa}{1992}]%
        {dj-main-paper}
\bibfield{author}{\bibinfo{person}{David Deutsch} {and}
  \bibinfo{person}{Richard Jozsa}.} \bibinfo{year}{1992}\natexlab{}.
\newblock \showarticletitle{Rapid solution of problems by quantum computation}.
\newblock \bibinfo{journal}{\emph{Proceedings of the Royal Society of London.
  Series A: Mathematical and Physical Sciences}} \bibinfo{volume}{439},
  \bibinfo{number}{1907} (\bibinfo{year}{1992}), \bibinfo{pages}{553--558}.
\newblock
\urldef\tempurl%
\url{https://doi.org/10.1098/rspa.1992.0167}
\showDOI{\tempurl}
\showeprint{https://royalsocietypublishing.org/doi/pdf/10.1098/rspa.1992.0167}


\bibitem[\protect\citeauthoryear{Gokhale, Baker, Duckering, Brown, Brown, and
  Chong}{Gokhale et~al\mbox{.}}{2019}]%
        {gokhale2019}
\bibfield{author}{\bibinfo{person}{Pranav Gokhale}, \bibinfo{person}{Jonathan~M
  Baker}, \bibinfo{person}{Casey Duckering}, \bibinfo{person}{Natalie~C Brown},
  \bibinfo{person}{Kenneth~R Brown}, {and} \bibinfo{person}{Frederic~T Chong}.}
  \bibinfo{year}{2019}\natexlab{}.
\newblock \showarticletitle{Asymptotic improvements to quantum circuits via
  qutrits}. In \bibinfo{booktitle}{\emph{Proceedings of the 46th International
  Symposium on Computer Architecture}}. \bibinfo{pages}{554--566}.
\newblock


\bibitem[\protect\citeauthoryear{Gottesman}{Gottesman}{1998}]%
        {gottesman1998}
\bibfield{author}{\bibinfo{person}{Daniel Gottesman}.}
  \bibinfo{year}{1998}\natexlab{}.
\newblock \showarticletitle{The Heisenberg representation of quantum
  computers}.
\newblock \bibinfo{journal}{\emph{arXiv preprint quant-ph/9807006}}
  (\bibinfo{year}{1998}).
\newblock


\bibitem[\protect\citeauthoryear{Lanyon, Jurcevic, Zwerger, Hempel, Martinez,
  D\"ur, Briegel, Blatt, and Roos}{Lanyon et~al\mbox{.}}{2013}]%
        {trapped-ions}
\bibfield{author}{\bibinfo{person}{B.~P. Lanyon}, \bibinfo{person}{P.
  Jurcevic}, \bibinfo{person}{M. Zwerger}, \bibinfo{person}{C. Hempel},
  \bibinfo{person}{E.~A. Martinez}, \bibinfo{person}{W. D\"ur},
  \bibinfo{person}{H.~J. Briegel}, \bibinfo{person}{R. Blatt}, {and}
  \bibinfo{person}{C.~F. Roos}.} \bibinfo{year}{2013}\natexlab{}.
\newblock \showarticletitle{Measurement-Based Quantum Computation with Trapped
  Ions}.
\newblock \bibinfo{journal}{\emph{Phys. Rev. Lett.}}  \bibinfo{volume}{111}
  (\bibinfo{date}{Nov} \bibinfo{year}{2013}), \bibinfo{pages}{210501}.
\newblock
Issue 21.
\urldef\tempurl%
\url{https://doi.org/10.1103/PhysRevLett.111.210501}
\showDOI{\tempurl}


\bibitem[\protect\citeauthoryear{Monroe, Raussendorf, Ruthven, Brown, Maunz,
  Duan, and Kim}{Monroe et~al\mbox{.}}{2014}]%
        {monroe2014}
\bibfield{author}{\bibinfo{person}{C Monroe}, \bibinfo{person}{R Raussendorf},
  \bibinfo{person}{A Ruthven}, \bibinfo{person}{KR Brown}, \bibinfo{person}{P
  Maunz}, \bibinfo{person}{L-M Duan}, {and} \bibinfo{person}{J Kim}.}
  \bibinfo{year}{2014}\natexlab{}.
\newblock \showarticletitle{Large-scale modular quantum-computer architecture
  with atomic memory and photonic interconnects}.
\newblock \bibinfo{journal}{\emph{Physical Review A}} \bibinfo{volume}{89},
  \bibinfo{number}{2} (\bibinfo{year}{2014}), \bibinfo{pages}{022317}.
\newblock


\bibitem[\protect\citeauthoryear{Nielsen}{Nielsen}{2006}]%
        {Nielsen_2006}
\bibfield{author}{\bibinfo{person}{Michael~A. Nielsen}.}
  \bibinfo{year}{2006}\natexlab{}.
\newblock \showarticletitle{Cluster-state quantum computation}.
\newblock \bibinfo{journal}{\emph{Reports on Mathematical Physics}}
  \bibinfo{volume}{57}, \bibinfo{number}{1} (\bibinfo{date}{feb}
  \bibinfo{year}{2006}), \bibinfo{pages}{147--161}.
\newblock
\urldef\tempurl%
\url{https://doi.org/10.1016/s0034-4877(06)80014-5}
\showDOI{\tempurl}


\bibitem[\protect\citeauthoryear{Paolini, Roversi, and Zorzi}{Paolini
  et~al\mbox{.}}{2019}]%
        {iqu}
\bibfield{author}{\bibinfo{person}{Luca Paolini}, \bibinfo{person}{Luca
  Roversi}, {and} \bibinfo{person}{Margherita Zorzi}.}
  \bibinfo{year}{2019}\natexlab{}.
\newblock \showarticletitle{Quantum Programming Made Easy}.
\newblock \bibinfo{journal}{\emph{Electronic Proceedings in Theoretical
  Computer Science}}  \bibinfo{volume}{292} (\bibinfo{date}{apr}
  \bibinfo{year}{2019}), \bibinfo{pages}{133--147}.
\newblock
\urldef\tempurl%
\url{https://doi.org/10.4204/eptcs.292.8}
\showDOI{\tempurl}


\bibitem[\protect\citeauthoryear{Quantum}{Quantum}{2021}]%
        {ibmq}
\bibfield{author}{\bibinfo{person}{IBM Quantum}.}
  \bibinfo{year}{2021}\natexlab{}.
\newblock
\newblock
\urldef\tempurl%
\url{https://quantum-computing.ibm.com/}
\showURL{%
\tempurl}


\bibitem[\protect\citeauthoryear{Raussendorf, Browne, and Briegel}{Raussendorf
  et~al\mbox{.}}{2003}]%
        {Raussendorf_2003}
\bibfield{author}{\bibinfo{person}{Robert Raussendorf},
  \bibinfo{person}{Daniel~E. Browne}, {and} \bibinfo{person}{Hans~J. Briegel}.}
  \bibinfo{year}{2003}\natexlab{}.
\newblock \showarticletitle{Measurement-based quantum computation on cluster
  states}.
\newblock \bibinfo{journal}{\emph{Physical Review A}} \bibinfo{volume}{68},
  \bibinfo{number}{2} (\bibinfo{date}{aug} \bibinfo{year}{2003}).
\newblock
\urldef\tempurl%
\url{https://doi.org/10.1103/physreva.68.022312}
\showDOI{\tempurl}


\bibitem[\protect\citeauthoryear{Svore, Roetteler, Geller, Troyer, Azariah,
  Granade, Heim, Kliuchnikov, Mykhailova, and Paz}{Svore et~al\mbox{.}}{2018}]%
        {q-sharp}
\bibfield{author}{\bibinfo{person}{Krysta Svore}, \bibinfo{person}{Martin
  Roetteler}, \bibinfo{person}{Alan Geller}, \bibinfo{person}{Matthias Troyer},
  \bibinfo{person}{John Azariah}, \bibinfo{person}{Christopher Granade},
  \bibinfo{person}{Bettina Heim}, \bibinfo{person}{Vadym Kliuchnikov},
  \bibinfo{person}{Mariia Mykhailova}, {and} \bibinfo{person}{Andres Paz}.}
  \bibinfo{year}{2018}\natexlab{}.
\newblock \showarticletitle{Q{\#}}. In \bibinfo{booktitle}{\emph{Proceedings of
  the Real World Domain Specific Languages Workshop 2018 on - {RWDSL}2018}}.
  \bibinfo{publisher}{{ACM} Press}.
\newblock
\urldef\tempurl%
\url{https://doi.org/10.1145/3183895.3183901}
\showDOI{\tempurl}


\bibitem[\protect\citeauthoryear{Tame, Bell, Di~Franco, Wadsworth, and
  Rarity}{Tame et~al\mbox{.}}{2014}]%
        {simons-experiment}
\bibfield{author}{\bibinfo{person}{M.~S. Tame}, \bibinfo{person}{B.~A. Bell},
  \bibinfo{person}{C. Di~Franco}, \bibinfo{person}{W.~J. Wadsworth}, {and}
  \bibinfo{person}{J.~G. Rarity}.} \bibinfo{year}{2014}\natexlab{}.
\newblock \showarticletitle{Experimental Realization of a One-Way Quantum
  Computer Algorithm Solving Simon's Problem}.
\newblock \bibinfo{journal}{\emph{Phys. Rev. Lett.}}  \bibinfo{volume}{113}
  (\bibinfo{date}{Nov} \bibinfo{year}{2014}), \bibinfo{pages}{200501}.
\newblock
Issue 20.
\urldef\tempurl%
\url{https://doi.org/10.1103/PhysRevLett.113.200501}
\showDOI{\tempurl}


\bibitem[\protect\citeauthoryear{Tame and Kim}{Tame and Kim}{2010}]%
        {bv-dj-tame}
\bibfield{author}{\bibinfo{person}{M.~S. Tame} {and} \bibinfo{person}{M.~S.
  Kim}.} \bibinfo{year}{2010}\natexlab{}.
\newblock \showarticletitle{Scalable method for demonstrating the Deutsch-Jozsa
  and Bernstein-Vazirani algorithms using cluster states}.
\newblock \bibinfo{journal}{\emph{Phys. Rev. A}}  \bibinfo{volume}{82}
  (\bibinfo{date}{Sep} \bibinfo{year}{2010}), \bibinfo{pages}{030305}.
\newblock
Issue 3.
\urldef\tempurl%
\url{https://doi.org/10.1103/PhysRevA.82.030305}
\showDOI{\tempurl}


\bibitem[\protect\citeauthoryear{Vallone, Donati, Bruno, Chiuri, and
  Mataloni}{Vallone et~al\mbox{.}}{2010}]%
        {deutschjozsa2010}
\bibfield{author}{\bibinfo{person}{Giuseppe Vallone}, \bibinfo{person}{Gaia
  Donati}, \bibinfo{person}{Natalia Bruno}, \bibinfo{person}{Andrea Chiuri},
  {and} \bibinfo{person}{Paolo Mataloni}.} \bibinfo{year}{2010}\natexlab{}.
\newblock \showarticletitle{Experimental realization of the Deutsch-Jozsa
  algorithm with a six-qubit cluster state}.
\newblock \bibinfo{journal}{\emph{Physical Review A}} \bibinfo{volume}{81},
  \bibinfo{number}{5} (\bibinfo{date}{May} \bibinfo{year}{2010}).
\newblock
\showISSN{1094-1622}
\urldef\tempurl%
\url{https://doi.org/10.1103/physreva.81.050302}
\showDOI{\tempurl}


\bibitem[\protect\citeauthoryear{van~de Wetering}{van~de Wetering}{2020}]%
        {vandewetering2020zxcalculus}
\bibfield{author}{\bibinfo{person}{John van~de Wetering}.}
  \bibinfo{year}{2020}\natexlab{}.
\newblock \bibinfo{title}{ZX-calculus for the working quantum computer
  scientist}.
\newblock
\newblock
\showeprint[arxiv]{2012.13966}~[quant-ph]


\bibitem[\protect\citeauthoryear{Walther, Resch, Rudolph, Schenck, Weinfurter,
  Vedral, Aspelmeyer, and Zeilinger}{Walther et~al\mbox{.}}{2005}]%
        {Walther_2005}
\bibfield{author}{\bibinfo{person}{P. Walther}, \bibinfo{person}{K.~J. Resch},
  \bibinfo{person}{T. Rudolph}, \bibinfo{person}{E. Schenck},
  \bibinfo{person}{H. Weinfurter}, \bibinfo{person}{V. Vedral},
  \bibinfo{person}{M. Aspelmeyer}, {and} \bibinfo{person}{A. Zeilinger}.}
  \bibinfo{year}{2005}\natexlab{}.
\newblock \showarticletitle{Experimental one-way quantum computing}.
\newblock \bibinfo{journal}{\emph{Nature}} \bibinfo{volume}{434},
  \bibinfo{number}{7030} (\bibinfo{date}{mar} \bibinfo{year}{2005}),
  \bibinfo{pages}{169--176}.
\newblock
\urldef\tempurl%
\url{https://doi.org/10.1038/nature03347}
\showDOI{\tempurl}


\end{thebibliography}

\begin{appendix}

\section{Cluster States}\label{app:clusters}

\begin{figure}[H]
  \scalebox{0.9}{
  \begin{minipage}{0.99\textwidth}
    
  \begin{flushleft}
    Cluster 1: ``Linear-3''
  \end{flushleft}

  \begin{minipage}{0.48\textwidth}
    \centering
    \begin{tikzpicture}  
      [scale=.9,auto=left,every node/.style={circle,fill=blue!20}]
        
      \node (a1) at (1,1) {$q_1$};  
      \node (a2) at (3,1) {$q_2$};  
      \node (a3) at (5,1) {$q_3$};
      
      \draw (a1) -- (a2); 
      \draw (a2) -- (a3);

    \end{tikzpicture}
  \end{minipage}\hfill
  \begin{minipage}{0.48\textwidth}
    {

      \def\gAxA{\op{R_z(-\alpha)}\w\A{gAxA}}
      \def\gBxA{\op{R_x(-\beta)}\w\A{gBxA}}


      \def\bA{\qv{q_{3}}{q_1}}


      \xymatrix@R=5pt@C=10pt{
          \bA & \gAxA &\gBxA &\n
      %
      %
      }
    }
  \end{minipage}
  \begin{flushright}
    $J(\alpha)(q_1, q_2); J(\beta)(q_2,q_3)$
  \end{flushright}
  \vspace{0mm}

  \begin{flushleft}
    Cluster 2: ``Linear-4''
  \end{flushleft}
  \begin{minipage}{0.40\textwidth}
    \centering
    \centering
    \begin{tikzpicture}  
      [scale=.9,auto=center,every node/.style={circle,fill=blue!20}]
        
      \node (a1) at (1,1) {$q_1$};  
      \node (a2) at (2.5,1) {$q_2$};  
      \node (a3) at (4,1) {$q_3$};
      \node (a4) at (5.5,1) {$q_4$};
      
      \draw (a1) -- (a2); 
      \draw (a2) -- (a3);
      \draw (a3) -- (a4);

    \end{tikzpicture}  
  \end{minipage}\hfill
  \begin{minipage}{0.40\textwidth}
    {
      \def\gAxA{\op{R_z(-\alpha)}\w\A{gAxA}}
      \def\gBxA{\op{R_x(-\beta)}\w\A{gBxA}}
      \def\gCxA{\op{R_z(-\gamma)}\w\A{gCxA}}
      \def\gDxA{\op{H}\w\A{gDxA}}


      \def\bA{\qv{q_{4}}{q_1}}


      \xymatrix@R=5pt@C=10pt{
          \bA & \gAxA &\gBxA &\gCxA &\gDxA &\n
      %
      %
      }
    }
  \end{minipage}
  \begin{flushright}
    $J(\alpha)(q_1, q_2); J(\beta)(q_2,q_3); J(\gamma)(q_3,q_4)$
  \end{flushright}
  \vspace{0mm}

  \begin{flushleft}
    Cluster 3: ``Horseshoe''
  \end{flushleft}
  \begin{minipage}{0.48\textwidth}
    \centering
    \begin{tikzpicture}  
      [scale=.9,auto=center,every node/.style={circle,fill=blue!20}]
        
      \node (a1) at (1,3) {$q_1$};
      \node (a2) at (1,1) {$q_2$};  

      \node (a3) at (3,3) {$q_3$};
      \node (a4) at (3,1) {$q_4$};  
      
      \draw (a1) -- (a2); 
      \draw (a1) -- (a3);
      \draw (a2) -- (a4);

    \end{tikzpicture} 
  \end{minipage}\hfill
  \begin{minipage}{0.48\textwidth}
    {
      \def\gAxA{\b\w\A{gAxA}}
      \def\gAxB{\b\w\A{gAxB}}
      \def\gBxA{\op{R_z(-\alpha)}\w\A{gBxA}}
      \def\gBxB{\op{R_z(-\beta)}\w\A{gBxB}}
      \def\gCxA{\op{H}\w\A{gCxA}}
      \def\gCxB{\op{H}\w\A{gCxB}}
      
      
      \def\bA{\qv{q_{3}}{q_1}}
      \def\bB{\qv{q_{4}}{q_2}}
      
      
      \xymatrix@R=5pt@C=10pt{
          \bA & \gAxA &\gBxA &\gCxA &\n
      \\  \bB & \gAxB &\gBxB &\gCxB &\n
      %
      %
      \ar@{-}"gAxA";"gAxB"
      }
    }
  \end{minipage}
  \begin{flushright}
    $CZ(q_1,q_2); J(\alpha)(q_1, q_3); J(\beta)(q_2,q_4)$
  \end{flushright}
  \vspace{0mm}

  \begin{flushleft}
    Cluster 4: ``Reverse Horseshoe''
  \end{flushleft}
  \begin{minipage}{0.48\textwidth}
    \centering
    \begin{tikzpicture}  
      [scale=.9,auto=center,every node/.style={circle,fill=blue!20}]
        
      \node (a1) at (1,3) {$q_1$};
      \node (a2) at (1,1) {$q_2$};  

      \node (a3) at (3,3) {$q_3$};
      \node (a4) at (3,1) {$q_4$};  
      
      \draw (a3) -- (a4); 
      \draw (a1) -- (a3);
      \draw (a2) -- (a4);

    \end{tikzpicture} 
  \end{minipage}\hfill
  \begin{minipage}{0.48\textwidth}
    {
      \def\gAxA{\op{R_z(-\alpha)}\w\A{gAxA}}
      \def\gAxB{\op{R_z(-\beta)}\w\A{gAxB}}
      \def\gBxA{\op{H}\w\A{gBxA}}
      \def\gBxB{\op{H}\w\A{gBxB}}
      \def\gCxA{\b\w\A{gCxA}}
      \def\gCxB{\b\w\A{gCxB}}
      
      
      \def\bA{\qv{q_{3}}{q_1}}
      \def\bB{\qv{q_{4}}{q_2}}
      
      
      \xymatrix@R=5pt@C=10pt{
          \bA & \gAxA &\gBxA &\gCxA &\n
      \\  \bB & \gAxB &\gBxB &\gCxB &\n
      %
      %
      \ar@{-}"gCxA";"gCxB"
      }
    }
  \end{minipage}
  \begin{flushright}
    $J(\alpha)(q_1, q_3); J(\beta)(q_2,q_4); CZ(q_1,q_2)$
  \end{flushright}
  \vspace{0mm}

  \begin{flushleft}
    Cluster 5: ``Box''
  \end{flushleft}
  \begin{minipage}{0.48\textwidth}
    \centering
    \begin{tikzpicture}  
      [scale=.9,auto=center,every node/.style={circle,fill=blue!20}]
        
      \node (a1) at (1,3) {$q_1$};
      \node (a2) at (1,1) {$q_2$};  

      \node (a3) at (3,3) {$q_3$};
      \node (a4) at (3,1) {$q_4$};  
      
      \draw (a1) -- (a2);
      \draw (a3) -- (a4); 
      \draw (a1) -- (a3);
      \draw (a2) -- (a4);

    \end{tikzpicture} 
  \end{minipage}\hfill
  \begin{minipage}{0.48\textwidth}
    {
      \def\gAxA{\b\w\A{gAxA}}
      \def\gAxB{\b\w\A{gAxB}}
      \def\gBxA{\op{R_z(-\alpha)}\w\A{gBxA}}
      \def\gBxB{\op{R_z(-\beta)}\w\A{gBxB}}
      \def\gCxA{\op{H}\w\A{gCxA}}
      \def\gCxB{\op{H}\w\A{gCxB}}
      \def\gDxA{\b\w\A{gDxA}}
      \def\gDxB{\b\w\A{gDxB}}
      
      
      \def\bA{\qv{q_{3}}{q_1}}
      \def\bB{\qv{q_{4}}{q_2}}
      
      
      \xymatrix@R=5pt@C=10pt{
          \bA & \gAxA &\gBxA &\gCxA &\gDxA &\n
      \\  \bB & \gAxB &\gBxB &\gCxB &\gDxB &\n
      %
      %
      \ar@{-}"gAxA";"gAxB"
      \ar@{-}"gDxA";"gDxB"
      }
    }
  \end{minipage}
  \begin{flushright}
    $CZ(q_1,q_2); J(\alpha)(q_1, q_3); J(\beta)(q_2,q_4); CZ(q_1,q_2)$
  \end{flushright}
  \end{minipage}
  }
  \vspace{0mm}

  \caption{Some Basic Cluster Conversions. The cluster is shown on the left with the equivalent
  circuit shown top right and measurement calculus pattern shown bottom right. Based on work by [Walther et al., 2005]}
  \label{fig:clusters}
\end{figure}

\section{Side by side code comparison}\label{app:codecmp}

\subsection{MCBeth program (OCaml)}
  
\begin{lstlisting}[language=Caml]
    open Backend.Types;;
    open Backend.Utils;;
    
    (** Build the cluster for quantum teleportation *)
    let teleport = standardize [
        Input(0);
        PrepList([1, 2]);
        J(0.0, 0, 1); 
        J(0.0, 1, 2);
    ];;
    
\end{lstlisting}
  
\subsection{Qiskit Program (Gate-based)}
  
\begin{lstlisting}[language=python]
   # Import qiskit and necessary methods/classes
    from qiskit import QuantumCircuit, QuantumRegister, ClassicalRegister

    # Setup the circuit
    quantum_reg = QuantumRegister(3, name="qreg")
    z_classical_reg = QuantumRegister(1, name="zcreg")
    x_classical_reg =  QuantumRegister(1, name="xcreg")
    circuit = QuantumCircuit(quantum_reg, z_classical_reg, x_classical_reg)
    
    # Define the steps of the teleportation
    def create_bell_pair(circuit, a, b)
        circuit.h(a)
        circuit.cx(a, b)
    
    def first_ops(circuit, state, a):
        circuit.cx(state, a)
        circuit.h(state)
        
    def measure_and_send(circuit, a, b):
        # Measures qubits a & b, 'sending' the results to b
        circuit.measure(a, 0)
        circuit.measure(b, 1)
        
    def rebuild_state(circuit, qubit, crz, crx):
        # Apply gates if the registers are in the state '1'
        qc.x(qubit).c_if(crx, 1)
        qc.z(qubit).c_if(crz, 1)
        
    # Build the teleportation circuit
    create_bell_pair(circuit, 1, 2)
    first_ops(circuit, 0, 1)
    measure_and_send(circuit, 0, 1)
    rebuild_state(circuit, 2, z_classical_reg, x_classical_reg)
\end{lstlisting}
  
\subsection{Qiskit Program (Measurement to Gate Based Translation)}
  
\begin{lstlisting}[language=python]
    # Import qiskit and necessary methods/classes
    from qiskit import QuantumCircuit, QuantumRegister, ClassicalRegister
    
    # Setup the circuit
    quantum_reg = QuantumRegister(3, name="qreg")
    z_classical_reg = QuantumRegister(1, name="zcreg") 
    x_classical_reg = QuantumRegister(1, name="xcreg")
    circuit = QuantumCircuit(quantum_reg, z_classical_reg, x_classical_reg)
    
    # Define the steps of the teleportation
    def entangle_pair(circuit, a, b)
        circuit.h(b)
        circuit.cx(a, b)
        circuit.h(b)
        
    def measure_and_send(circuit, a, b):
        # Measures qubits a & b, 'sending' the results to b
        circuit.measure(a, 0)
        circuit.measure(b, 1)
        
    def rebuild_state(circuit, qubit, crz, crx):
        # Apply gates if the registers are in the state '1'
        qc.x(qubit).c_if(crx, 1)
        qc.z(qubit).c_if(crz, 1)
        
    # Build the teleportation circuit
    # Programmer would need to define the input state here.
    entangle(circuit, 0, 1)
    entangle(circuit, 1, 2)
    measure_and_send(circuit, 0, 1)
    rebuild_state(circuit, 2, z_classical_reg, x_classical_reg)
\end{lstlisting}
  

\end{appendix}

\end{document}